\documentclass[10pt,twocolumn,reqno,amsmath,amssymb,aps,prb]{revtex4-2}

\usepackage{graphicx}
\usepackage[justification=justified, singlelinecheck=false]{caption}
\usepackage[justification=justified]{subcaption}

\usepackage{hyperref}    
\usepackage{xcolor}
\usepackage{leftindex}

\usepackage{dcolumn}
\usepackage{bm}

\begin{document}

\preprint{APS/123-QED}

\title{Spin waves in the bilayer van der Waals magnet CrSBr}%

\author{Rob den Teuling$^1$}
\altaffiliation[Corresponding author:]{ r.denteuling@tudelft.nl.}

\author{Ritesh Das$^1$}
\author{Artem V. Bondarenko$^{1,2}$}

\author{\\ Elena V. Tartakovskaya$^{1,2,3}$}

\author{Gerrit E. W. Bauer$^{4,5}$}

\author{Yaroslav M. Blanter$^1$ \vspace{3mm}}

\affiliation{${}^1$ \hspace{-3mm} Kavli Institute of Nanoscience, Delft University of Technology, Lorentzweg 1, 2628 CJ, Delft, The Netherlands, \vspace{2mm}}%

\affiliation{${}^2$ \hspace{-3mm} V.G. Baryakhtar Institute of Magnetism of the NAS of Ukraine, 36b Vernadsky Boulevard, 03142 Kiev, Ukraine,\hspace{-3mm} \vspace{2mm}}

\affiliation{${}^3$ \hspace{-3mm} Institute of Spintronics and Quantum Information, Faculty of Physics and Astronomy, Adam Mickiewicz University, Poznań, Uniwersytetu Poznańskiego 2, 61-614 Poznań, Poland,\hspace{-2.5mm} \vspace{2mm}}

\affiliation{${}^4$ \hspace{-3mm} WPI-AIMR \& IMR  \& CSIS, Tohoku University, 2-1-1 Katahira, Sendai 980-8577, Japan, 
\vspace{2mm}}

\affiliation{${}^5$ \hspace{-3mm} Kavli Institute for Theoretical Sciences, University of the Chinese Academy of Sciences, Beijing 10090, China. \vspace{2mm}}

\date{\today}

\begin{abstract}
We derive analytical expressions for the spin wave frequencies and precession amplitudes in monolayer and antiferromagnetically coupled bilayer CrSBr under in-plane external magnetic fields. The analysis covers the antiferromagnetic, ferromagnetic, and canted phases, demonstrating that the spin wave frequencies in all phases are tunable by the applied magnetic field. We discuss the roles of intra- and interlayer exchange interactions, triaxial anisotropy, and intralayer dynamic dipolar fields in controlling the magnetization dynamics.
\end{abstract}

\maketitle


\section{\label{sec:level1}Introduction}

Since the discovery of long-range magnetic ordering in CrI\textsubscript{3} \cite{Huang_2017} and Cr\textsubscript{2}Ge\textsubscript{2}Te\textsubscript{6} \cite{8083602} in 2017, two-dimensional (2D) van der Waals (vdW) magnets have been considered promising for applications such as spintronic logic circuits \cite{ohmic}. These devices employ spin waves, or magnons, as information carriers that do not suffer from Ohmic losses inherent to electron charge transport. For a comprehensive review on spin-wave computing, see Ref. \cite{mahmoud}. Subsequently, numerous 2D vdW magnetic materials have been identified. For example, Fe\textsubscript{3}GeTe\textsubscript{2} \cite{mat1} and VSe\textsubscript{2} \cite{mat2} remain magnetic down to single atomic layers. Others have been discovered experimentally and/or predicted theoretically \cite{structure}. They include ferromagnets, antiferromagnets, and layered antiferromagnets \cite{gibertini}. Magnetism in these materials has been shown to be tunable through various external stimuli, such as electric fields \cite{electric, electric2}, pressure \cite{pressure, pressure2}, strain \cite{strain, Makars}, and doping \cite{doping, doping2}. Furthermore, these materials display multiple magnetic phase transitions that are sensitive to external parameters. This versatility makes them strong candidates for flexible, low-power applications across various electronic and quantum computing platforms.

Magnetic order in the layered 2D vdW material CrSBr was predicted by Guo et al. \cite{theory}, and has  been confirmed  by experiments, see e.g. Ref  \cite{experiments}.  CrSBr has a tetragonal crystal lattice, unlike most 2D vdW magnets that are hexagonal. It is the first layered material found to exhibit in-plane ferromagnetic (FM) order in monolayers but antiferromagnetic (AFM) coupling between the layers, similar to synthetic antiferromagnets (SAFs) \cite{synthetic2}. The AFM order with vanishing net magnetization is robust against weak magnetic stray fields \cite{emergent}. The FM ordering in the layers and relatively weak AFM coupling leads to low spin wave frequencies in the GHz regime. Mono- and bilayers of CrSBr are ideal for studying the excitation, manipulation, and detection of 2D magnons. The magnetic properties of monolayer \cite{screening} and bilayer \cite{triax} CrSBr, including effects of biaxial \cite{Cho} and triaxial \cite{Cham} anisotropies, have been reported. A comprehensive study of the magnon dispersion and precession amplitudes under intra- and interlayer coupling, triaxial anisotropy, and dynamic dipolar interactions, is still lacking.

Here we present analytical expressions for the spin wave frequencies and precession amplitudes as functions of intra- and interlayer exchange coupling, triaxial anisotropy constant, dynamic dipolar interactions, and in-plane external fields. The paper is organized as follows. \hyperref[II]{Sec. II} develops the theoretical framework. We start from Hamiltonians for mono- and bilayers to derive effective fields for the linearized Landau-Lifshitz (LL) equation. In \hyperref[III]{Sec. III} and \hyperref[IV]{Sec. IV}, we provide the frequencies and eigenvectors of spin waves in the mono- and bilayer, respectively,  as a function of in-plane external magnetic fields along the principal axes.  \hyperref[V]{Sec. V} summarizes the results.

\section{\label{II}Theoretical Framework}

\subsection{\label{sec:level2}Hamiltonian}

\begin{figure} [htp]

\subfloat[]{%
  \includegraphics[clip,width=0.76\columnwidth]{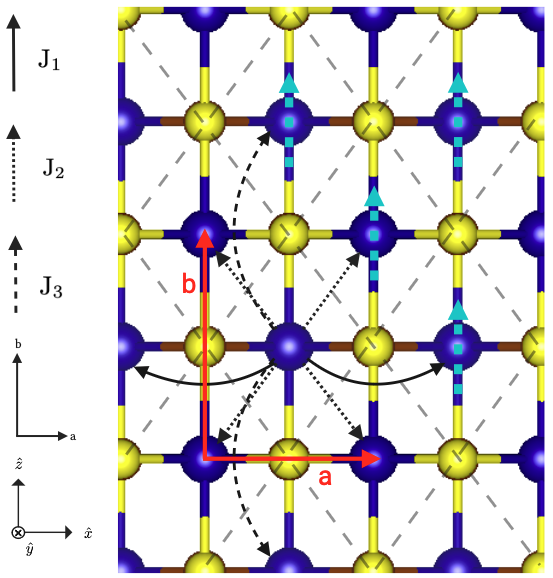}%
}

\subfloat[]{%
  \includegraphics[clip,width=0.75\columnwidth]{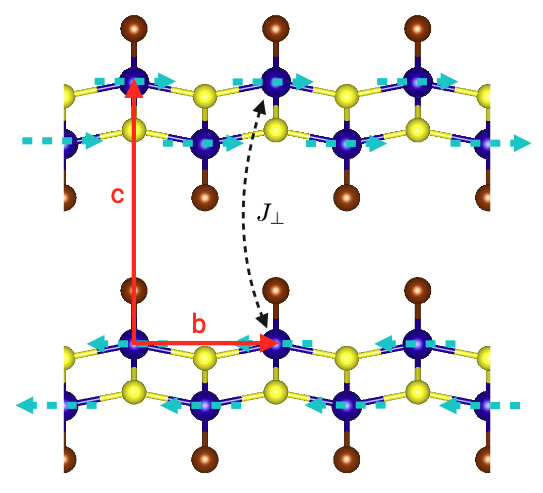}%
}

\caption{ \small Top (a) and side (b) views of bilayer CrSBr, illustrating the axes, lattice vectors, and exchange coupling constants \(J_i\). Thin dashed lines indicate the primitive unit cell. The thick dashed arrows illustrate the magnetic moment orientations and AFM order. Chromium atoms are represented in blue, sulfur in yellow, and bromine in brown.}
\label{fig1}
\end{figure}

\hyperref[fig1]{FIG. 1} shows a bilayer of CrSBr with a tetragonal unit cell and AA stacking along the \(y\)-direction \cite{stack}.

\noindent
The magnetic moments of the Cr atoms in a monolayer lie in the \(ab\)-plane. In a multilayer, the spins order in CrSBr in an A-type AFM ordering, in which ferromagnetic monolayers align along the \(b\)-axis with alternating orientations. Bulk CrSBr has a Néel temperature of 137 K \cite{bulk}, while the monolayer is ferromagnetic below 146 K \cite{tempe}. In CrSBr the dipolar field appears to be of the same order of magnitude as the magnetocrystalline anisotropy \cite{dipolar}.

CrSBr is predicted to exhibit a triaxial anisotropy that favors spins to lie in the plane with a favored \(b\)-direction, with strong intralayer FM coupling and weak interlayer AFM coupling \cite{marg}. Three coupling strengths dominate the intralayer exchange, viz. \(\mathcal{J}_3 >\mathcal{J}_1 >\mathcal{J}_2 > 0\), see \hyperref[fig1]{FIG. 1}, while the AFM interlayer exchange coupling \(\mathcal{J}_{\perp} < 0 \). The intralayer coupling constants of the two atoms in the unit cell (\hyperref[appA]{Appendix A}) are not necessarily the same \cite{bulk} but are assumed here to be equal. This simplifies the analytical treatment since the unit cell contains effectively only one chromium spin. The Brillouin zone is then twice as large without optical intralayer modes. We discuss this approximation in \hyperref[appA]{Appendix A} for the monolayer and \hyperref[appB]{Appendix B} for the bilayer.

Parameter values for CrSBr vary across the literature \cite{values1, values2, bulk, theory}. \hyperref[tab1]{Table I} summarizes the numerical values used in this work. We adopt the parameters for the monolayer from density functional theory (DFT). The intralayer exchange coupling constants are taken from Ref. \cite{strain} and the magnetic anisotropy constants from Ref. \cite{triax}, where the sign of the anisotropy constants is changed to remain consistent with our choice of Hamiltonian. For the interlayer exchange coupling, we prefer the experimental value \( \mathcal{J}_{\perp}=-15.5 \mu eV \) \cite{Boix_Constant_2023} with a modulus that is an order of magnitude larger than the estimate from DFT \cite{triax}. Both the spin-orbit coupling (SOC) and shape anisotropy originating from dipole-dipole interactions contribute to the total magnetic anisotropy of monolayer CrSBr \cite{triax}. In the ground state, the second term dominates. Finally, the saturation magnetization of the CrSBr monolayer is \(M_s=0.48\) T \cite{gibertini, Ms2}.

\vspace{-1mm}

\begin{table}[h]
\caption{\label{tab1}%
Numerical values of the in-plane lattice constants \emph{a} and \emph{b}, the exchange constants \(\mathcal{J}_i\) \cite{strain}, and the anisotropy constants \(\mathcal{D}_i\) \cite{triax}.
}
\begin{ruledtabular}
\begin{tabular}{ccc}
    $\emph{a}$ & 3.54 & $10^{-10} \mathrm{m}$ \\ 
    $\emph{b}$ & 4.73 & $10^{-10}\mathrm{m}$ \\ \hline
    $\mathcal{J}_1$ & 3.54 & $\mathrm{meV}$ \\ 
    $\mathcal{J}_2$ & 3.08 & $\mathrm{meV}$ \\
    $\mathcal{J}_3$ & 4.15 & $\mathrm{meV}$ \\ 
    $\mathcal{J}_{\perp}$ & -15.5 & $\mu$$\mathrm{eV}$ \\ \hline 
   $ \mathcal{D}_x$ & -12 & $\mu$$\mathrm{eV}$ \\
    $\mathcal{D}_y$ & -78 & $\mu$$\mathrm{eV}$ \\
    $\mathcal{D}_z$ & 0 & $\mu$$\mathrm{eV}$ \\ 
\end{tabular}
\end{ruledtabular}
\end{table}
\footnotetext[1]{Here's the first, from Ref.~\onlinecite{feyn54}.}
\footnotetext[2]{Here's the second.}
\footnotetext[3]{Here's the third.}
\footnotetext[4]{Here's the fourth.}
\footnotetext[5]{And etc.} 

\noindent

In the Hamiltonian for a bilayer
\begin{equation} \label{1}
    H_{Bi} = {\hspace{-0.5mm}}H_A +{\hspace{-0.5mm}}H_B + H_{int}.
\end{equation}

\noindent
\(H_A\) and \(H_B\) are the Hamiltonians for the isolated monolayers A and B, respectively. They include the Zeeman energy from the external field \(\Vec{B}_0\) (\(H_{Ext}\)), the intralayer exchange coupling (\(H_{Ex}\)), and magnetic anisotropy (\(H_{An}\)) \cite{rezende}. Without dipolar interactions (see below) 
\begin{equation} \label{2}
    \begin{aligned}
        {\hspace{-0.5mm}}H_A = & \underbrace{-\sum_{j}\gamma \hbar \Vec{B}_0\cdot{\hspace{-0.5mm}}\Vec{S}_{j,A}}_{H_{Ext}} \underbrace{-\sum_{j,\sigma} \mathcal{J}_{\sigma}{\hspace{-0.5mm}}\Vec{S}_{j,A} \cdot {\hspace{-0.5mm}}\Vec{S}_{j+\sigma,A}}_{H_{Ex}} \\
        & \underbrace{-\sum_{j} \Big[\mathcal{D}_x\hspace{1mm} {\hspace{-0.5mm}}S_{x,j,A}^2 + \mathcal{D}_y\hspace{1mm} {\hspace{-0.5mm}}S_{y,j,A}^2 + \mathcal{D}_z\hspace{1mm} {\hspace{-0.5mm}}S_{z,j,A}^2 \Big]}_{H_{An}}, \\
    \end{aligned}
\end{equation}

\noindent
where \(j\) counts the local spins and \(\sigma\) sums over the nearest neighbors of the \(j\)-th spin, \(\gamma\) is the gyromagnetic ratio, \(g\) the Landé g-factor, and \(\mu_B\) the Bohr magneton. The intralayer exchange coupling constants are denoted by \(\mathcal{J}_{\sigma}\) as in \hyperref[fig1]{FIG. 1}, with 
\begin{equation} \label{3}
\mathcal{J}_{\pm \Vec{a}}=\mathcal{J}_1, \ 
\mathcal{J}_{\pm \Vec{d}_1}  = \mathcal{J}_{\pm \Vec{d}_2}=\mathcal{J}_2, \ 
\mathcal{J}_{\pm \Vec{b}}=\mathcal{J}_3, 
\end{equation}

\noindent
where \(\Vec{d}_1 = \frac{1}{2}(\Vec{a} + \Vec{b})\) and \(\Vec{d}_2 = \frac{1}{2}(\Vec{a} - \Vec{b})\). The Hamiltonian of layer B is written by making the substitution \(A \rightleftharpoons B\).
\noindent
The exchange interaction between the layers
\begin{equation} \label{4}
    H_{int} = -\sum_{j} \mathcal{J}_{\perp}\hspace{-0.5mm}\Vec{S}_{j,A} \cdot \hspace{-0.5mm}\Vec{S}_{j,B}.
\end{equation}

\noindent
is antiferromagnetic since \(J_{\perp}<0\).

We address the spin dynamics by making a classical approximation that allows us to write the equations of motion of the magnetization as a torque or LL equation,
\begin{equation} \label{5}
    \frac{d \vec{m}_j(t)}{dt} = -\gamma \left [ \vec{m}_j(t) \times \vec{B}_{eff} \right ],
\end{equation}

\noindent
where \( \Vec{m}_j (t) = \gamma \hbar \Vec{S}_j /M_s \), the normalized magnetic moment at a lattice site j on layer A or B, and \(\Vec{B}_{eff}\) the effective magnetic field that is the functional derivative \cite{stancil}, 
\begin{equation} \label{6}
    \Vec{B}_{eff} = - \frac{1}{M_s} \frac{\partial \varepsilon}{\partial \Vec{m}_l},
\end{equation}

\noindent
of the energy density per unit cell \(\varepsilon\) that follows from the Hamiltonian by replacing the quantum spin operators by classical magnetization amplitudes via \(\Vec{S} = \Vec{M}/(\gamma \hbar)\):
\begin{equation} \label{7}
    \begin{aligned}
        \varepsilon = &- M_s \sum_{l,\sigma} \frac{\mathcal{J}_{\sigma} S}{\gamma \hbar} \Vec{m}_l\cdot\Vec{m}_{l,\sigma} \\
        & -M_s \sum_{l}\left(\frac{\mathcal{D}_x S}{\gamma \hbar}m_{x,l}^2 - \frac{\mathcal{D}_y S}{\gamma \hbar}m_{y,l}^2 - \frac{\mathcal{D}_z S}{\gamma \hbar}m_{z,l}^2\right) \\
        & - M_s \sum_{l}\Vec{B}_0 \cdot \Vec{m}_l, \\
    \end{aligned}
\end{equation}

\noindent
where \(l\) runs over all sites in the unit cell and \(S = M_s/(\gamma \hbar)\). We linearize the LL equation by the spin wave Ansatz
\begin{equation} \label{8}
    \begin{aligned}
        \Vec{m}_l & = m_z\hat{z} + (m_{x}\hat{x} + m_{y}\hat{y}) e^{i(\omega t - \Vec{k}\cdot\Vec{r}_l)} \\
        & \equiv m_z\hat{z} + \underline{m}_{x}\hat{x} + \underline{m}_{y}\hat{y} \ .
    \end{aligned}
\end{equation}

\noindent
where \( \underline{m}_{x,y} = m_{x,y} e^{i(\omega t - \Vec{k}\cdot\Vec{r}_l)}\),  in the small parameters \(m_x, m_y \ll m_z=1\).

The exchange coupling terms in the Hamiltonian lead to an effective field of the form \(\Vec{B}_{Ex} = \alpha_i m_z\hat{z} + \beta_i\underline{m}_x\hat{x} + \beta_i\underline{m}_y\hat{y} \). Defining \( J_k \equiv (\mathcal{J}_k S)/ (\gamma \hbar) \), these terms read
\begin{equation} \label{9}
    \begin{aligned}
        & \alpha_1 = 2(J_1 + J_3), \ \alpha_2 = 4J_2, \\
        & \beta_1 = 2[J_1\cos(k_xa) + J_3\cos(k_zb)], \\
        & \beta_2 = 2J_2 \left[\cos \left(\frac{1}{2}k_xa + \frac{1}{2}k_zb \right) + \cos \left( \frac{1}{2}k_xa - \frac{1}{2}k_zb \right) \right]\hspace{-1.2mm}.
    \end{aligned}
\end{equation}
\noindent
The anisotropy field reads
\begin{equation} \label{10}
    B_{An,\eta} = 2D_{\eta} \underline{m}_{\eta}, D_{\eta} \equiv \mathcal{D}_{\eta} S/(\gamma \hbar), \ \eta = x, y, z.
\end{equation}
 
\subsection{\label{sec:level2}Dipolar Fields}

The dipolar interaction can be categorized into intralayer and interlayer contributions. Stamps \cite{Stamps} demonstrated that the dynamic dipolar \emph{interlayer} interactions decay exponentially with distance. Furthermore, as we will show elsewhere \cite{Teuling}, in the case of bilayer CrSBr the product of the film thickness \(d\) and the wave vector \(k\) are numerically small and can be disregarded. 

The continuum approximation significantly simplifies an analytical treatment of dipole-dipole interaction (Ewald)  sums \cite{intradipolar}. It is valid in the long-wave limit (small \(kd\)), and sufficiently accurate for larger \(k\) at which the exchange energy dominates. The tetragonal distortion induces a very small in-plane anisotropy in the static dipolar interaction energy \cite{triax} whose effect on the effective field we disregard.

The dipolar field in a thin film with in-plane magnetization along the \(z\)-axis then reads \cite{D5, D4}
\begin{equation} \label{11}
    \Vec{B}_{dip} = - \mu_0 M_s f(k) m_y \hat{y} - \mu_0 M_s \frac{k_x^2}{k^2}(1-f(k))m_x \hat{x},
\end{equation}

\noindent
where
\begin{equation} \label{12}
    f(k) = \frac{1-e^{-kd}}{kd}.
\end{equation}

The thickness \(d\) of a van der Waals atomic monolayer is strictly speaking not well defined since it depends on the chemical bonds between the Cr, S, and Br atoms  \hyperref[fig1]{FIG. 1},. However, its main role is to regulate a divergence in the dipolar Ewald sum in strictly two dimensions. The numerical results are not very sensitive to deviations from our choice \(d = b/2\). In the following, we add the intralayer dipolar fields  \hyperref[11]{Eq. (11)} to \(\Vec{B}_{eff}\).

\begin{figure*} \label{fig2}
\centering
 \begin{center}
     \begin{subfigure}{0.48\textwidth} \label{2a}
    \includegraphics[width=\linewidth]{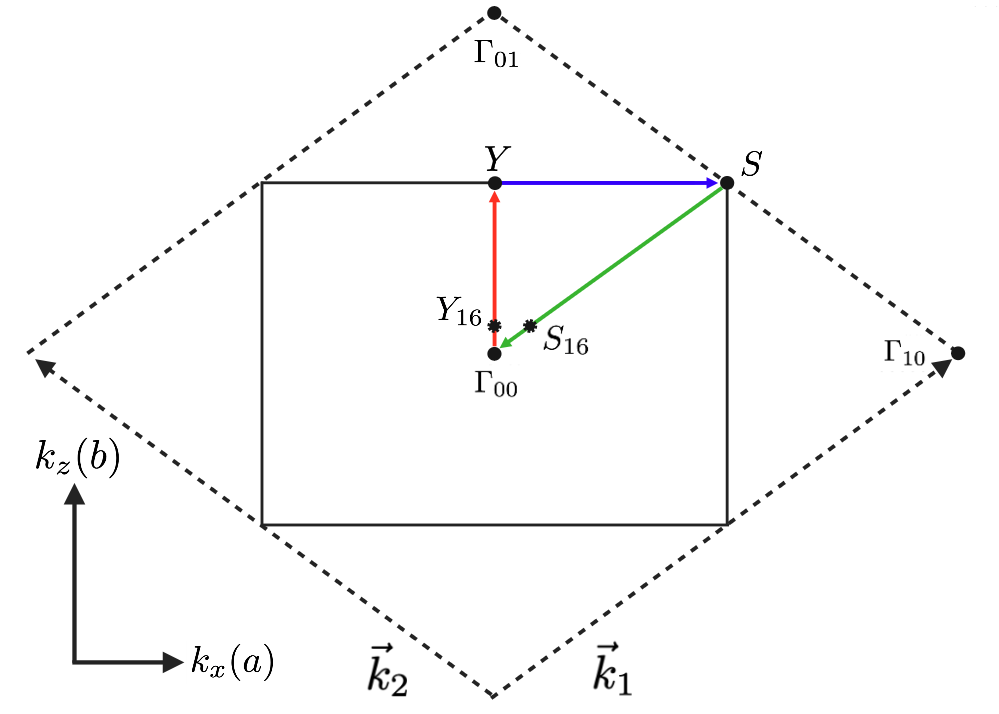}
    \vspace{-7mm}
    \caption{} 
  \end{subfigure}
 \begin{subfigure}{0.48\textwidth} \label{2b}
    \includegraphics[width=\linewidth]{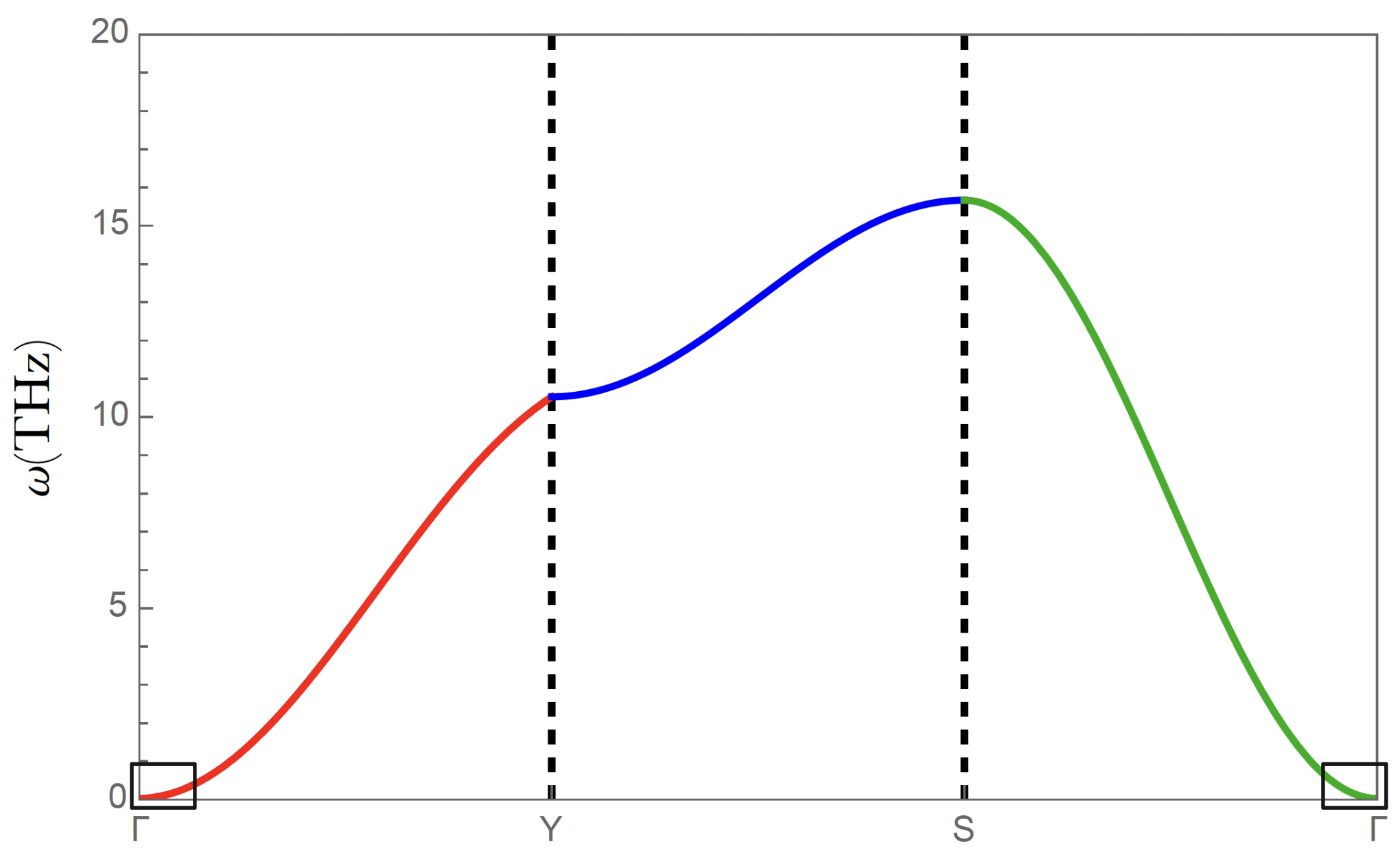}
    \vspace{-7mm}
    \caption{} 
  \end{subfigure}
    \begin{subfigure}{0.48\textwidth} \label{2c}
    \includegraphics[width=\linewidth]{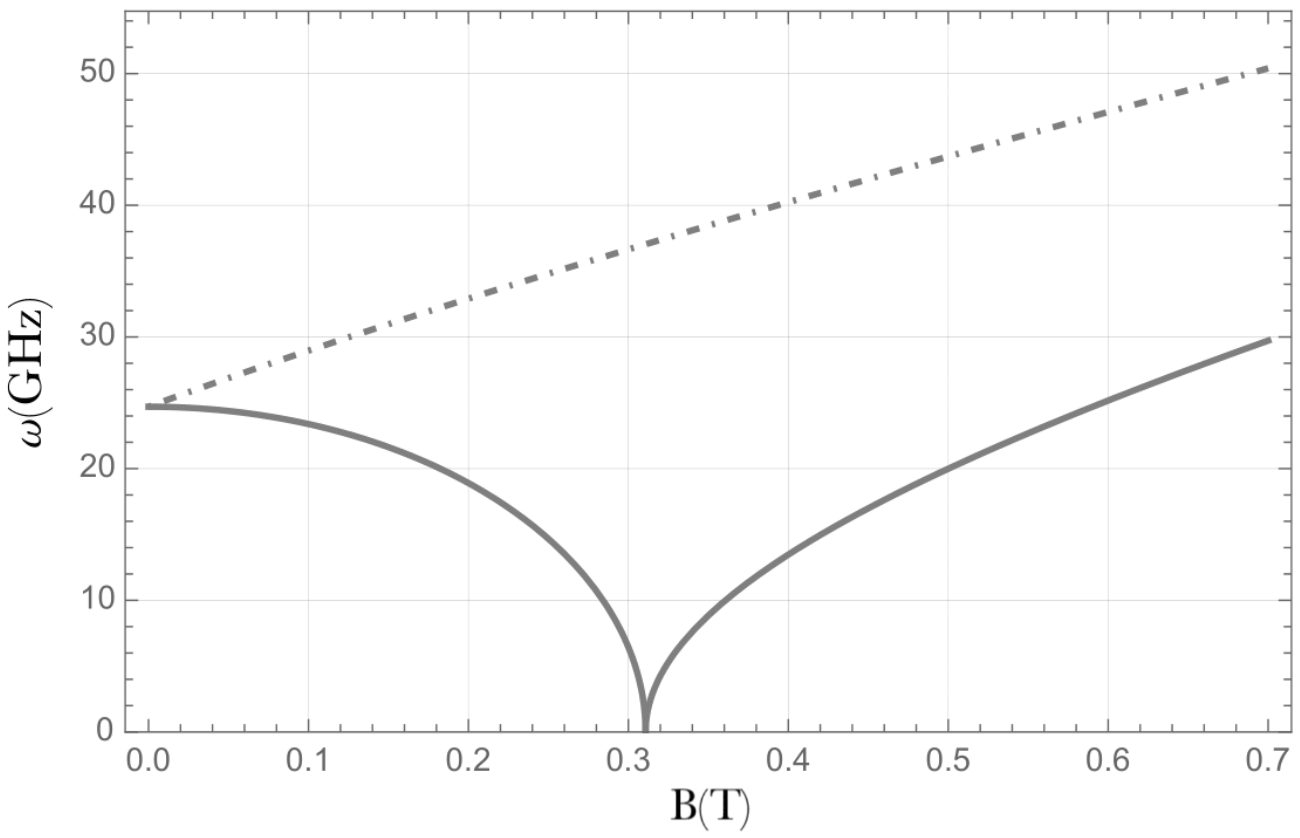}
    \vspace{-7mm}
    \caption{} 
  \end{subfigure}
 \begin{subfigure}{0.48\textwidth} \label{2d}
    \includegraphics[width=\linewidth]{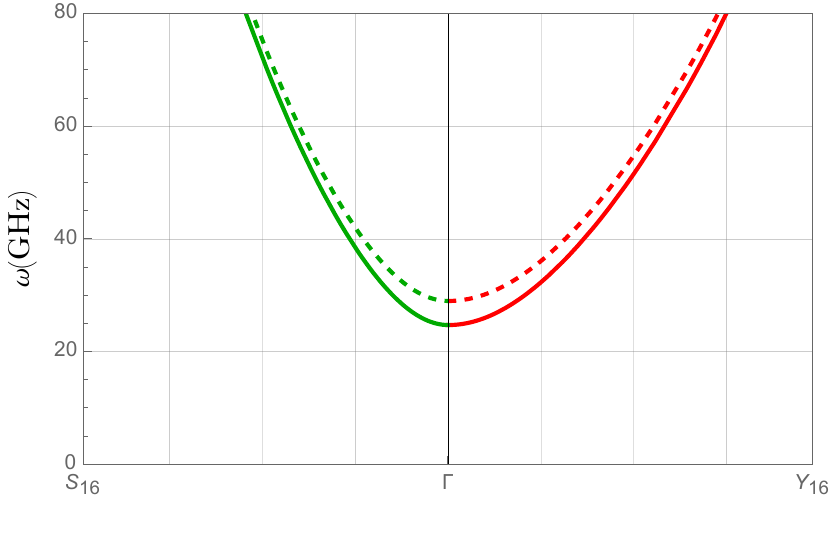}
    \vspace{-7mm}
    \caption{} 
  \end{subfigure}
    \begin{subfigure}{0.48\textwidth} \label{2e}
    \includegraphics[width=\linewidth]{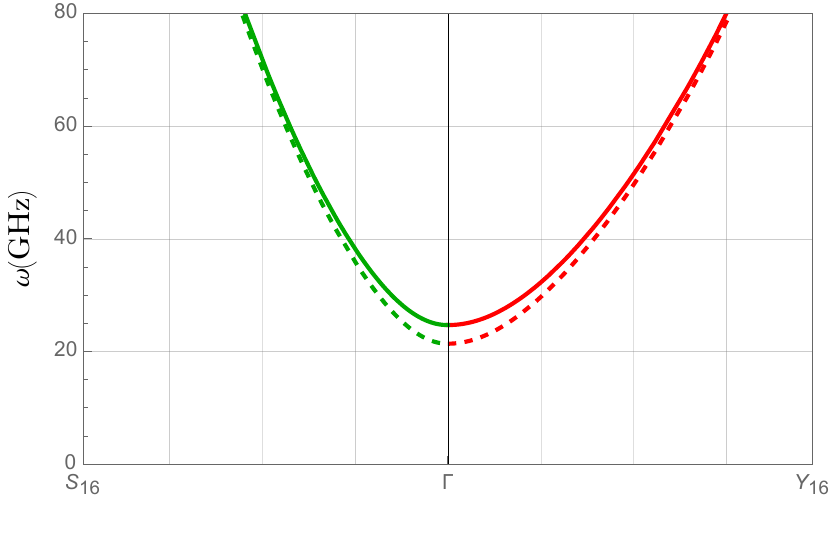}
    \vspace{-7mm}
    \caption{} 
  \end{subfigure}
 \begin{subfigure}{0.48\textwidth} \label{2f}
    \includegraphics[width=\linewidth]{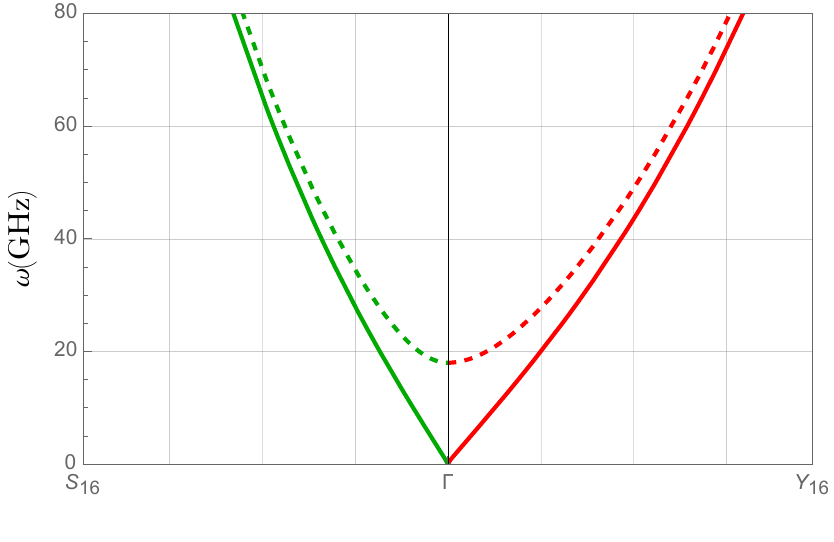}
    \vspace{-7mm}
    \caption{} 
  \end{subfigure}
  \end{center}
    \vspace{-3mm}
 \caption{\small \textbf{(a)} Brillouin zone (BZ) of CrSBr. The solid lines correspond to two inequivalent chromium atoms per unit cell, while the dashed lines indicate  the BZ when the two chromium atoms are approximated to be equivalent. \textbf{(b)} Dispersion of monolayer CrSBr without applied magnetic field. The colors of the curve indicate the path in the BZ, while the small rectangles indicate the dispersions around the origin that are enlarged in (d-f). \textbf{(c)} Monolayer resonance frequencies (\(k=0\)). The dash-dotted curve corresponds to an external magnetic field along the easy axis, and the solid curve is for a field along the intermediate axis (in-plane normal to the easy axis). \textbf{(d-f) }Monolayer dispersion relations close to the \(\Gamma\)-point for two magnetic fields (for higher fields shown as dotted curves):\textbf{ (d)} External fields of 0 and \(-\frac{1}{2}J_{\perp}\) along the easy axis. \textbf{(e)} External fields of 0 and \(\frac{1}{2}B_{sat}\) along the intermediate axis. \textbf{(f)} External fields of \(B_{sat}\) and \(\frac{3}{2} B_{sat}\) along the intermediate axis. \(B_{sat}\) is the critical field that fully aligns all spins.}
 \end{figure*}

\section{\label{III}Monolayer}
Our results for the equilibrium state and FMR frequency of the monolayer agree with the existing literature \cite{rezende, Cho, Cham}. Here, we focus on the equilibrium and excited magnetization with an external magnetic field applied along the \(a\)- and \(b\)-axes. Substituting the effective field \hyperref[7]{Eq. (7)}  into \hyperref[5]{Eq. (5)} leads to a \(2\times2\) eigenvalue problem in the \([m_x, m_y]^T\) basis.

When the external field is applied along the easy axis, the frequency dispersion of the lowest mode across the  Brillouin zone (BZ) is
\begin{equation} \label{13}
    \omega_- = \frac{\gamma}{2\pi} \sqrt{AB},
\end{equation}

\noindent
where
\begin{equation} \label{14}
    \begin{aligned}
        A & = B_0 + 2(D_z - D_y) + [\alpha_1 - \beta_1] + [\alpha_2 - \beta_2] \\
        & + \mu_0 M_s f(k) \\
        B & = B_0 + 2(D_z - D_x) + [\alpha_1 - \beta_1] + [\alpha_2 - \beta_2] \\
        & + \mu_0 M_s \frac{k_x^2}{k^2}(1-f(k)).
    \end{aligned}
\end{equation}

\noindent
The non-normalized eigenvector reads
\begin{equation} \label{15}
    \Vec{m}_{\omega_-} = \begin{bmatrix}
    1 \\
    -i\sqrt{\frac{B}{A}}
    \end{bmatrix}\hspace{-1.2mm}.
\end{equation}

A magnetic field along the intermediate axis causes a canting of the sublattice spins in the (a)-direction by an angle  $\theta = \arcsin [B_0/(2(D_z-D_x)]$ from the easy axis. In the canted basis introduced in \hyperref[appC]{Appendix C} the exchange field reads \(\Vec{B}_{Ex} = \alpha_i m_{\gamma}\hat{e}_{\gamma} + \beta_i \underline{m}_{\alpha}\hat{e}_{\alpha} + \beta_i \underline{m}_{\beta}\hat{e}_{\beta} \) and the dipolar field is \(\Vec{B}_{dip} = - \mu_0 M_s f(k) m_{\beta} \hat{e}_{\beta} - \mu_0 M_s \frac{k_{\alpha}^2}{k^2}(1-f(k))m_{\alpha} \hat{e}_{\alpha}\). The saturation field that fully aligns the magnetic moments along the intermediate axis is \( B_{sat} = 2(D_z - D_x) \). The frequency dispersion in the canted and saturated (\(\theta = \pi/2\)) phases is
\begin{equation} \label{16}
    \omega_- = \frac{\gamma}{2\pi} \sqrt{CD},
\end{equation}

\noindent
where
\begin{equation} \label{17}
    \begin{aligned}
        C & = B_0\sin(\theta) + 2(D_x \sin^2(\theta) + D_z \cos^2(\theta) - D_y) \\
        & + [\alpha_1 - \beta_1] + [\alpha_2 - \beta_2] + \mu_0 M_s f(k), \\
        D & = B_0\sin(\theta) + 2(D_z -D_x) \cos(2\theta) \\ 
        & + [\alpha_1 - \beta_1] + [\alpha_2 - \beta_2] + \mu_0 M_s \frac{k_{\alpha}^2}{k^2}(1-f(k)), \\
    \end{aligned}
\end{equation}

\noindent
with eigenvectors
\begin{equation} \label{18}
    \Vec{m}_{\omega_-}= \begin{bmatrix}
    1 \\
    -i\sqrt{\frac{D}{C}}
    \end{bmatrix}\hspace{-1.2mm}.
\end{equation}

\noindent
For fields applied along the intermediate axis, the saturation field at which the FMR frequency vanishes is \( B_{sat} \approx 0.31\) T.

\hyperref[2a]{FIG. 2a} show the BZ of monolayer CrSBr \cite{giant} spanned by the reciprocal unit vectors \(\Vec{k}_1 = [2\pi/a \hspace{3mm} 2\pi/b]^T\) and \(\Vec{k}_2 = [-2\pi/a \hspace{3mm} 2\pi/b]^T\). The colored arrows show the path in the BZ used in plotting the dispersion. \hyperref[2b]{FIG. 2b} shows the magnon dispersion of a CrSBr monolayer in the absence of external magnetic fields. The effects of the magnetic crystal anisotropy and  dipolar interactions are relatively small compared to that of the intralayer exchange, except for the small wave vector region close to the \(\Gamma\)-point, as shown in the boxes of \hyperref[2b]{FIG. 2b}. \hyperref[2c]{FIG. 2c} shows the resonance frequencies and excited modes for the collinear and canted phases as a function of the applied field. \hyperref[2d]{FIG. 2d-f} show the dispersion relations of the monolayer with external fields applied along the easy and intermediate axes. The red and green curves follow the \(\Gamma_{00} \rightarrow Y_{16}\) and \(\Gamma_{00} \rightarrow S_{16}\) paths, respectively, corresponding to \(1/16^{th}\) of the paths from \(\Gamma_{00}\) to the BZ boundary \hyperref[fig2]{FIG. 2a}. A field along the easy axis causes an upward shift of the entire spectrum with increasing field. A transverse field initially shifts the frequencies downwards until vanishing at \(B_{sat}\), beyond which it increases monotonically with field.
\vspace{7mm}

\section{\label{IV}Bilayer}

We now turn to the magnetic properties of bilayer CrSBr in the presence of an external magnetic field applied along the easy and intermediate axes. In the basis
\begin{equation} \label{19}
    \begin{bmatrix}
    \hspace{-0.5mm}m_{\alpha,A} \\
    \hspace{-0.5mm}m_{\alpha,B} \\
    \hspace{-0.5mm}m_{\beta,A} \\
    \hspace{-0.5mm}m_{\beta,B}
        \end{bmatrix}\hspace{-1.2mm}.
\end{equation}
\noindent
we have to solve an LL  \(4\times4\) matrix equation, where \( \alpha = x\) \( (\alpha = z)\) when magnetic moments are collinear with the easy (intermediate) axes,  respectively.

When external magnetic fields vanish, the ground state is an antiferromagnet with antiparallel  orientation of the monolayer magnetizations  \cite{callen}.  Since the interlayer exchange field is weak, an external field along the easy axis induces spin-flip transition to a collinear ferromagnetic state  at a critical field strength \(B_{crit}\), rather than a spin-flop transition observed in many other AFMs. In the following, we compute the equilibrium magnetization, frequency, and eigenvectors for an in-plane external magnetic field applied along the easy and intermediate axes. The FMR and antiferromagnetic resonance (AFMR) frequencies in bulk systems are known \cite{Cham, Cho}, the spin wave dispersion and eigenvector spectra of the bilayer including the triaxial anisotropy and dipolar interactions are to our knowledge new.

\subsection{Antiferromagnetic phase}

\hspace{\parindent} The resonance frequencies for the AFM phase are found from the eigenvalue equation
\begin{align} \label{20}
    \omega \begin{bmatrix}
    \hspace{-0.5mm}m_{x,A} \\
    \hspace{-0.5mm}m_{x,B} \\
    \hspace{-0.5mm}m_{y,A} \\
    \hspace{-0.5mm}m_{y,B}
        \end{bmatrix} = \gamma \begin{bmatrix}
0 & 0 & iE_1 & -iJ_{\perp} \\
0 & 0 & iJ_{\perp} & -iE_2 \\
-iF_1  & iJ_{\perp}  & 0 & 0  \\
-iJ_{\perp} & iF_2 & 0 & 0
\end{bmatrix}\begin{bmatrix}
    \hspace{-0.5mm}m_{x,A} \\
    \hspace{-0.5mm}m_{x,B} \\
    \hspace{-0.5mm}m_{y,A} \\
    \hspace{-0.5mm}m_{y,B}
        \end{bmatrix}\hspace{-1.2mm},
\end{align}

\noindent
where
\begin{equation} \label{21}
    \begin{aligned}
        E_1 & = B_0 + 2(D_z - D_y) + [\alpha_1 - \beta_1] + [\alpha_2 - \beta_2] \\
        & -J_{\perp} + \mu_0 M_s f(k), \\
        E_2 & = - B_0 + 2(D_z - D_y) + [\alpha_1 - \beta_1] + [\alpha_2 - \beta_2] \\
        & -J_{\perp} + \mu_0 M_s f(k), \\
        F_1 & = B_0 + 2(D_z - D_x) + [\alpha_1 - \beta_1] + [\alpha_2 - \beta_2] \\
        & -J_{\perp} + \mu_0 M_s \frac{k_x^2}{k^2}(1-f(k)), \\
        F_2 & = -B_0 + 2(D_z - D_x) + [\alpha_1 - \beta_1] + [\alpha_2 - \beta_2] \\
        & -J_{\perp} + \mu_0 M_s \frac{k_x^2}{k^2}(1-f(k)),
    \end{aligned}
\end{equation}

\noindent
The antisymmetry of the external field along the magnetic moments (terms \(E_1\) and \(E_2\), and \(F_1\) and \(F_2\)) renders the analytical forms of the frequencies complex:

\begin{figure*} \label{fig3} [h]
 \begin{center}

     \begin{subfigure}{0.48\textwidth} \label{3a}
    \includegraphics[width=\linewidth]{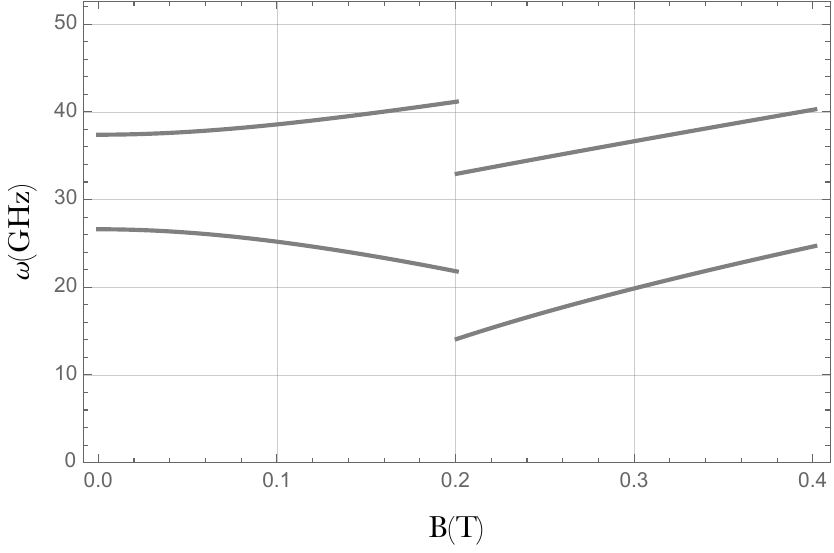}
    \vspace{-6mm}
    \caption{} 
  \end{subfigure}
 \begin{subfigure}{0.48\textwidth} \label{3b}
    \includegraphics[width=\linewidth]{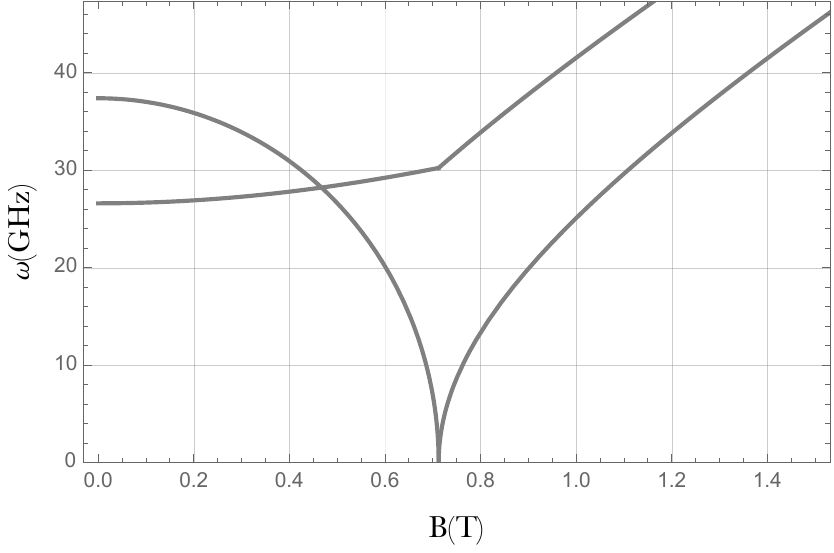}
    \vspace{-6mm}
    \caption{} 
  \end{subfigure}
  
    \begin{subfigure}{0.48\textwidth} \label{3c}
    \includegraphics[width=\linewidth]{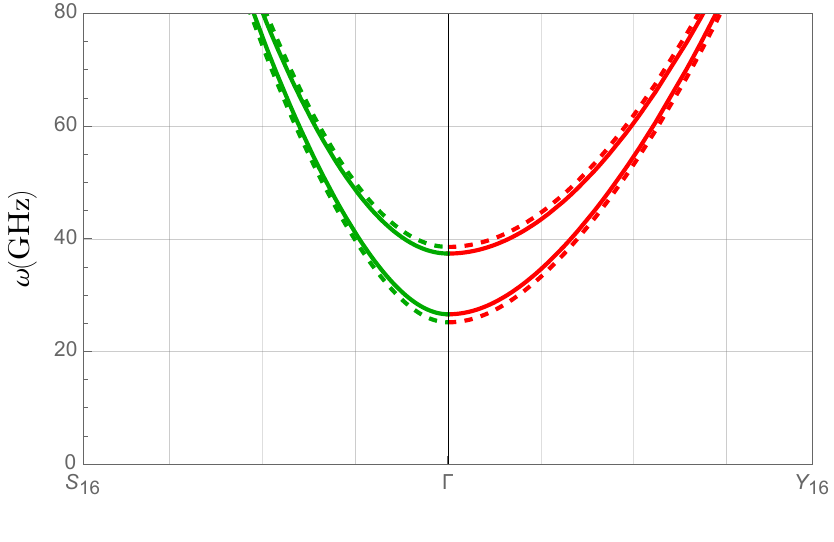}
    \vspace{-10mm}
    \caption{} 
  \end{subfigure}
 \begin{subfigure}{0.48\textwidth} \label{3d}
    \includegraphics[width=\linewidth]{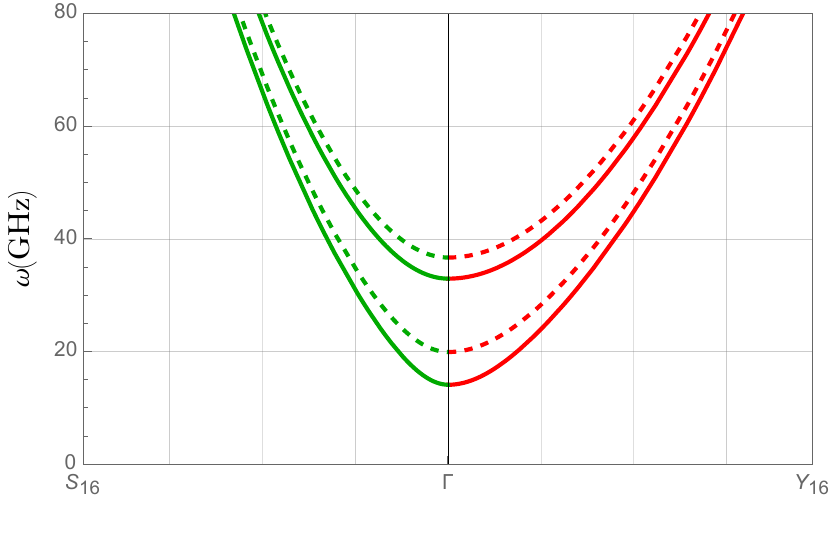}
    \vspace{-10mm}
    \caption{} 
  \end{subfigure}
  
    \begin{subfigure}{0.48\textwidth} \label{3e}
    \includegraphics[width=\linewidth]{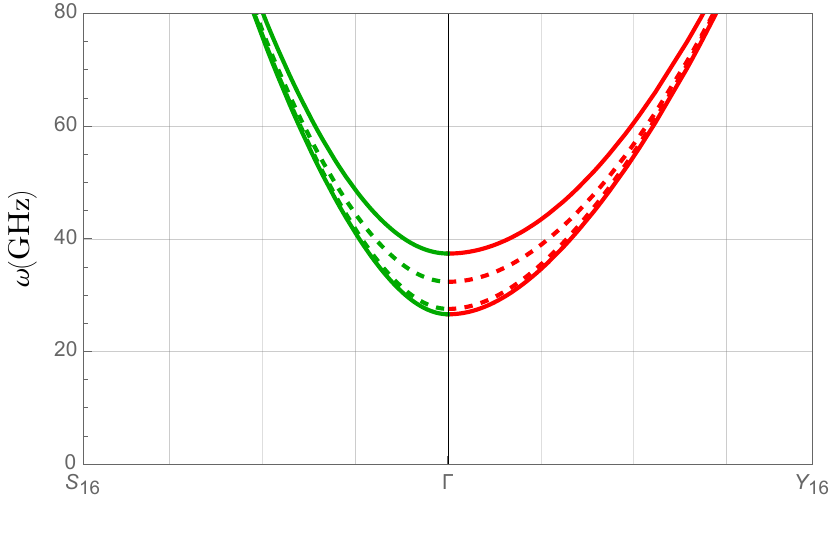}
    \vspace{-10mm}
    \caption{} 
  \end{subfigure}
 \begin{subfigure}{0.48\textwidth} \label{3f}
    \includegraphics[width=\linewidth]{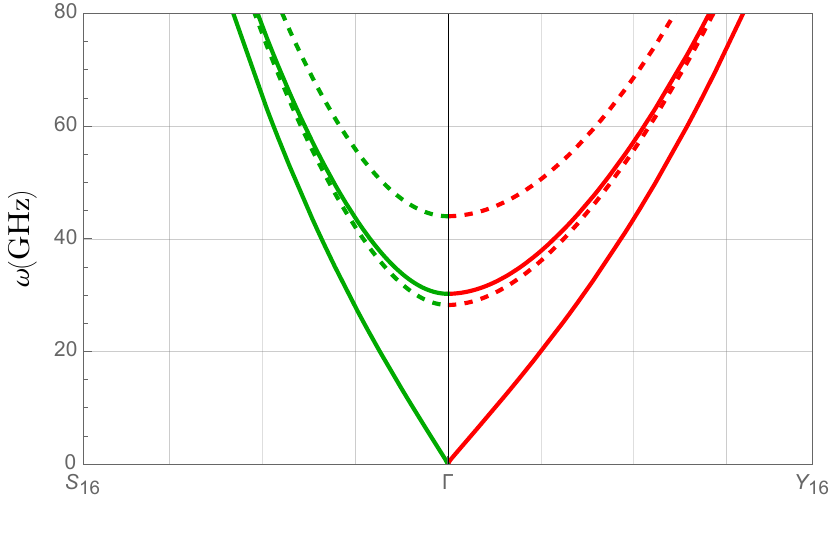}
    \vspace{-10mm}
    \caption{} 
  \end{subfigure}
  \end{center}
    \vspace{-5mm}
 \caption{\small \textbf{(a,b)} Bilayer magnetic resonance frequencies (\(k=0\)) as function of magnetic field: \textbf{(a)} AFM (left of the transition at \(B_0<0.2\) T) and FM ( \(B_0>0.2\) T) phases for a field direction along the easy axis. \textbf{(b)} Canted (\(B<0.71\) T) and saturated FM (\(B>0.71\) T) phases for an in-plane field normal to the easy axis.  \textbf{(c-f)} Bilayer dispersion relations close to the \(\Gamma\)-point for two magnetic field strengths (dotted curves indicate higher fields): \textbf{(c)} \(B_0=0\) and \(B_0=\frac{1}{2}B_{crit}^{flip}\) (dotted line) along the easy axis. \textbf{(d)} External fields of \(B_0=B_c^{flip}\) and \(B_0=\frac{3}{2}B_{crit}^{flip}\) along the easy axis, where \(B_{crit}^{flip}\) aligns the spins ferromagnetically along the easy axis. \textbf{(e)} External fields of \(B_0=\)o and \(B_0=\frac{1}{2}B_{sat}^{cant}\) along the intermediate axis. \textbf{(f)} External fields of \(B_{sat}^{cant}\) and \(\frac{3}{2}B_{sat}^{cant}\) along the intermediate axis.  \(B_{sat}^{cant}\) aligns the spins ferromagnetically along the intermediate in-plane normal-to-easy axis.}
 \end{figure*}
\begin{widetext}
\begin{equation} \label{22}
    \omega_{\mp} = \frac{\gamma}{2\pi} \frac{1}{\sqrt{2}} \sqrt{E_1F_1 + E_2F_2 - 2J_{\perp}^2 \mp \sqrt{\left( E_1F_1 + E_2F_2 - 2J_{\perp}^2 \right)^2 - 4\left( E_1E_2 - J_{\perp}^2\right)\left( F_1F_2 - J_{\perp}^2 \right)}}.
    \vspace{-5mm}
\end{equation}
\end{widetext}

The two positive frequency branches correspond to the lower frequency (\(\beta\)-mode) and higher frequency (\(\alpha\)-mode) excitations of the magnetic moments in the unit cell \cite{rezende}. When \(D_x\), \(D_y\), and the dipolar interaction vanish, we recover the well-known result by Kittel \cite{kittel}. 

\subsection{Spin-Flip Phase}

An external field along the easy axis induces a spin-flip transition from the AFM to FM configuration at a critical value \(B^{flip}_{crit}\), as illustrated in \hyperref[fig7]{FIG. 7} in \hyperref[appD]{Appendix D}. The equilibrium configuration minimizes the energy of \hyperref[7]{Eq. (7)} as a function of \(\theta_A\) and \(\theta_B\), where \(m_{x,i} = \cos\theta_i\) and \(m_{z,i} = \sin\theta_i\), \(i=A,B\), and \(\theta_i\) the angle with the \(x\)-axis. When \(B_0 < B_{crit}^{flip}\) the ground state is an AFM with \(\theta_A = -\theta_B = \pi/2 \), as discussed in Ref. \cite{callen}. When \(B_0 \geq B_{crit}^{flip}\)  $\theta_A = \theta_B = \pi/2$ minimizes the energy. By comparing the energies of AFM and FM states we find \( B^{flip}_{crit} = -J_{\perp} \approx\) \(0.2\) T \cite{Boix_Constant_2023}.
The frequencies of the two modes are
\begin{equation} \label{23}
    \omega_{\mp} = \frac{\gamma}{2\pi} \sqrt{(G \mp J_{\perp})(H \mp J_{\perp})},
\end{equation}

\noindent
where
\begin{equation} \label{24}
    \begin{aligned}
        G & = B_0 + 2(D_z - D_y) + [\alpha_1 - \beta_1]+ [\alpha_2 - \beta_2]\\
        &  + J_{\perp} + \mu_0 M_s f(k), \\
        H & = B_0 + 2(D_z - D_x) + [\alpha_1 - \beta_1]+ [\alpha_2 - \beta_2] \\
        &  + J_{\perp} + \mu_0 M_s \frac{k_x^2}{k^2}(1-f(k)), \\
    \end{aligned}
\end{equation}

\noindent
with eigenvectors
\begin{equation} \label{25}
    \begin{aligned}
        & \Vec{m}_{\omega_-}&= \begin{bmatrix}
        1 \\
        1 \\
        i\sqrt{\frac{H-J_{\perp}}{G-J_{\perp}}} \\
        i\sqrt{\frac{H-J_{\perp}}{G-J_{\perp}}}
        \end{bmatrix}\hspace{-1.2mm}, \hspace{3mm}\Vec{m}_{\omega_+} &= \begin{bmatrix}
        1 \\
        -1 \\
        i\sqrt{\frac{H+J_{\perp}}{G+J_{\perp}}} \\
        -i\sqrt{\frac{H+J_{\perp}}{G+J_{\perp}}} \\
        \end{bmatrix}\hspace{-1.2mm},
    \end{aligned}
\end{equation}

\noindent
As for the monolayer, we focus on the regime close to the \(\Gamma\)-point. \hyperref[3a]{FIG. 3a} displays the resonance frequencies of the AFM  and FM  phases in a bilayer. At the spin-flip transition, both frequencies drop abruptly. In the FM phase, both mode frequencies increase again with the external field \cite{rezende}.

\hyperref[3c]{FIG. 3c} and \hyperref[3d]{FIG. 3d} show the dispersion below and above the spin-flip transition, respectively. At \(B_0 = B_{crit}^{flip}\), the frequencies are minimal and shift to higher values again with increasing field.
\vspace{7mm}

\subsection{Canted Phase}

An external magnetic field applied along the intermediate axis (normal and in-plane to the easy axis) leads to a canting of the layer magnetizations as illustrated in \hyperref[fig8]{FIG. 8} in \hyperref[appD]{Appendix D} that becomes a ferromagnet at the saturation value \(B^{cant}_{sat}\). In the canted phase, the magnetic moments of the individual layers are non-collinear.  The LL equation can best be solved in sublattice-specific coordinate systems \hyperref[appC]{Appendix C}. The intralayer dipolar fields \hyperref[11]{Eq. (11)} in the respective bases are
\begin{equation} \label{26}
    \Vec{B}_{dip,i} = - \mu_0 M_s f(k) m_{\beta,i} \hat{e}^i_{\beta} - \mu_0 M_s \frac{k_{\alpha,i}^2}{k^2}(1-f(k))m_{\alpha,i} \hat{e}^i_{\alpha},
\end{equation}
\noindent
where \(i={A,B}\) indicates the sublattice. In subsequent expressions  \(k^2_{\alpha,i}\) can be written as a function of \(k_x\) and \(k_z\) via \hyperref[c2]{Eqs. (C2)} and \hyperref[c3]{(C3)}. Minimizing the energy yields the canting angle \(\sin \theta = B_0/[2(D_z - D_x - J_{\perp})] \). The resonance frequencies for the canted phase solve the LL equation
\begin{align} \label{27}
    \omega \begin{bmatrix}
    \hspace{-0.5mm}m_{\alpha,A} \\
    \hspace{-0.5mm}m_{\alpha,B} \\
    \hspace{-0.5mm}m_{\beta,A} \\
    \hspace{-0.5mm}m_{\beta,B}
        \end{bmatrix} = \gamma \begin{bmatrix}
0 & 0 & iI & -iJ_{\perp} \\
0 & 0 & -iJ_{\perp} & iI \\
-iK_A  & iK_1  & 0 & 0  \\
iK_1 & -iK_B & 0 & 0
\end{bmatrix}\begin{bmatrix}
    \hspace{-0.5mm}m_{\alpha,A} \\
    \hspace{-0.5mm}m_{\alpha,B} \\
    \hspace{-0.5mm}m_{\beta,A} \\
    \hspace{-0.5mm}m_{\beta,B}
        \end{bmatrix}\hspace{-1.2mm},
\end{align}

\noindent
where
\begin{equation} \label{28}
    \begin{aligned}
        I & = B_0\sin \theta + 2\left(D_x\sin^2\theta + D_z\cos^2\theta - D_y\right) \\
        & + [\alpha_1 - \beta_1] + [\alpha_2 - \beta_2]-J_{\perp}\cos(2\theta) + \mu_0 M_s f(k), \\
        K_A & = B_0\sin(\theta) + 2(D_z - D_x)\cos(2\theta) + [\alpha_1 - \beta_1] \\
        & + [\alpha_2 - \beta_2] -J_{\perp}\cos(2\theta) + \mu_0 M_s \frac{k_{\alpha,A}^2}{k^2}(1-f(k)), \\
        K_B & = B_0\sin(\theta) + 2(D_z - D_x)\cos(2\theta) + [\alpha_1 - \beta_1] \\
        & + [\alpha_2 - \beta_2] -J_{\perp}\cos(2\theta) + \mu_0 M_s \frac{k_{\alpha,B}^2}{k^2}(1-f(k)), \\
        K_1 & = -J_{\perp}\cos(2\theta) .
    \end{aligned}
\end{equation}

\noindent
The solutions read
\begin{widetext}
\begin{equation} \label{29}
    \omega_{\mp} = \frac{\gamma}{2\pi} \frac{1}{\sqrt{2}} \sqrt{I(K_A + K_B) + 2J_{\perp}E \mp \sqrt{\left( I(K_A + K_B)  + 2J_{\perp}E\right)^2 - 4\left(I^2 - J_{\perp}^2\right)\left( K_A K_B - E^2 \right)}}.
\end{equation}
\end{widetext}
The analytical solutions for the eigenfrequencies and eigenvectors are cumbersome again, this time due to \( k^2_{\alpha,A} \neq k^2_{\alpha,B}\) for  \(0 < \theta < \frac{\pi}{2}\) (terms \(K_A\) and \(K_B\)).  

At a saturation field \( B^{cant}_{sat} = 2(D_z - D_x - J_{\perp}) \approx 0.71\) T and above both layer magnetizations are parallel to the magnetic field direction. When \(B_0 \ge 2(D_z - D_x - J_{\perp})\)) we set \(\cos\theta=0\) and \(\sin\theta=1\),  leading to the frequencies
\begin{equation} \label{30}
        \omega_{\mp} = \frac{\gamma}{2\pi} \sqrt{(L \mp J_{\perp})(M \mp J_{\perp})},
\end{equation}

\noindent
where
\begin{equation} \label{31}
    \begin{aligned}
        L & = B_0 + 2(D_x - D_y)+ [\alpha_1 - \beta_1] + [\alpha_2 - \beta_2] \\
        &  + J_{\perp} + \mu_0 M_s f(k), \\
        M & = B_0 + 2(D_x - D_z)+ [\alpha_1 - \beta_1] + [\alpha_2 - \beta_2]\\
        &  + J_{\perp} + \mu_0 M_s \frac{k_z^2}{k^2}(1-f(k)),
    \end{aligned}
\end{equation}

\noindent
with eigenvectors
\begin{equation} \label{32}
    \begin{aligned}
        & \Vec{m}_{\omega_-} &= \begin{bmatrix}
        1 \\
        1 \\
        i\sqrt{\frac{M-J_{\perp}}{L-J_{\perp}}} \\
        i\sqrt{\frac{M-J_{\perp}}{L-J_{\perp}}}
        \end{bmatrix}\hspace{-1.2mm}, \hspace{3mm}\Vec{m}_{\omega_+} &= \begin{bmatrix}
        1 \\
        -1 \\
        i\sqrt{\frac{M+J_{\perp}}{L+J_{\perp}}} \\
        -i\sqrt{\frac{M+J_{\perp}}{L+J_{\perp}}} \\
        \end{bmatrix}\hspace{-1.2mm},
    \end{aligned}
\end{equation}

\noindent
in the basis \([m_{z,A}, m_{z,B}, m_{y,A}, m_{y,B}]^T\). This solution is identical to the spin-flip FM phase after substituting \(x \rightleftharpoons z\) . \hyperref[3b]{FIG. 3b} illustrates the resonance frequencies for the canted (below \(B^{cant}_{sat} \)) and saturated (above  \(B^{cant}_{sat} \)) phases of the bilayer. In the canted phase we may expect a magnon Hanle effect at the  mode crossing at \(B_0\approx0.47\) T \cite{Hanle}. \hyperref[3e]{FIG. 3e} pictures the dispersion in the canted phase below saturation. Both bands decrease in energy for an increasing external field. \hyperref[3f]{FIG. 3f} shows the dispersion in the saturated phase. At \(B_0 = B_{sat}^{cant}\), the lower band becomes soft, above which all frequencies increase monotonically with the applied field.

In all phases, for both the mono- and bilayer, inclusion of the dynamic dipolar field increases the frequency around the \(\Gamma\)-point by \(\sim11\%\).

\section{\label{V}Conclusion}

We derived spin wave frequencies and eigenvectors from the linearized monolayer and bilayer CrSBr under in-plane magnetic fields including FM intralayer and AFM interlayer exchange couplings, triaxial magnetic anisotropy, intralayer dipolar interactions, and external magnetic fields along the easy and intermediate axes.

In principle, our approach  applies to synthetic antiferromagnets and can be generalized to multilayer systems  at the cost of losing closed analytic forms.

Our results highlight the tunability of spin wave spectra in CrSBr through external magnetic fields and detail the interplay between the external field, inter- and intralayer exchange coupling, magnetic anisotropy, and dynamic dipolar interactions. They can be tested by magnetic resonance spectroscopy. They form the input to compute magnetic stray fields, propagating magnon spectroscopy,  Hanle effects, and diffuse spin transport including magnon spin conductivities, and spin Seebeck coefficients.

\begin{acknowledgments}
This publication is part of the project "Ronde Open Competitie XL" (file number OCENW.XL21.XL21.058) and "Ronde Open Competitie ENW pakket 21-3" (file number OCENW.M.21.215) which are (partly) financed by the Dutch Research Council (NWO). A.V.B. was supported by the EIC Pathfinder PALANTIRI project. E.V.T. was supported by the National Science Center of Poland, project no. UMO-2023/49/B/ST3/02920. G.B. was supported by JSPS Kakenhi Grants 22H04965 and JP24H02231. We thank Samer Kurdi and Fabian Gerritsma for insightful discussions. Images were made with BioRender and the crystallographic model was taken from VESTA.
\end{acknowledgments}

\appendix

\section{Two-sublattice monolayer CrSBr} \label{appA}

\begin{figure}[h] \label{fig4}
    \centering
    \includegraphics[scale=0.38]{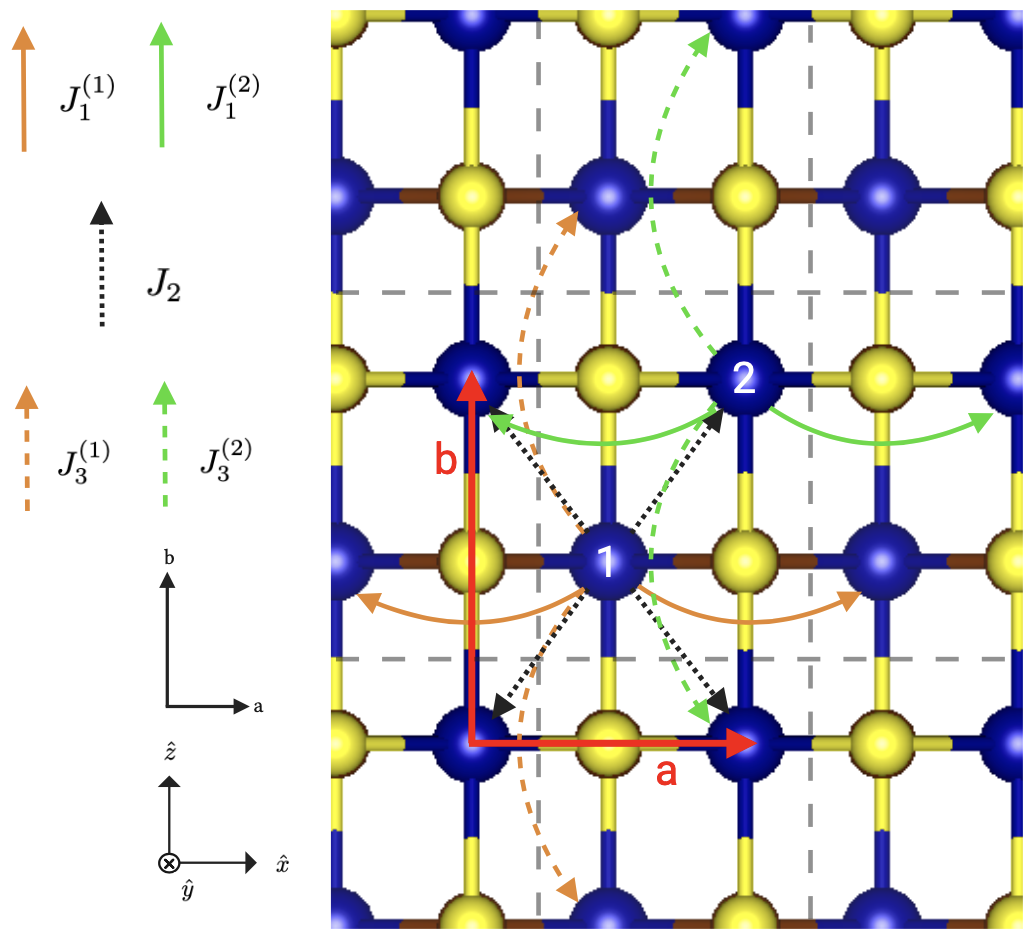}
    \caption{Anisotropic intralayer exchange coupling constants.}
\end{figure}

As discussed in the main text, the precise values of the intralayer exchange coupling parameters are not well known. Working with two slightly different couplings in the \textit{x}- and \textit{z}-directions could be better than the isotropic model used in the main text. \hyperref[fig4]{FIG. 4} illustrates the magnetic configuration with anisotropic exchange and two chromium atoms per unit cell. Compared to the isotropic exchange model the magnetic unit cell becomes twice as large. The number magnon modes in a half as large Brillouin zone doubles. The analytical diagonalization of the corresponding  \(4\times4\) eigenvalue matrices for the monolayer is tedious and we do not show the lengthy solutions here.

\noindent
The Hamiltonian for sublattice \(1\) in layer \(A\) for such a configuration reads
\begin{equation} \label{a1}
    \begin{aligned}
        {\hspace{-0.5mm}}\Hat{H}^{(1)}_A = & \underbrace{-\sum_{j,\sigma_1} \mathcal{J}_{\sigma_1}{\hspace{-0.5mm}}\Vec{S}_{j,A}^{(1)} \cdot {\hspace{-0.5mm}}\Vec{S}^{(1)}_{j+\sigma_1,A}}_{\Hat{H}_{Ex,1}} \\
        & \underbrace{-\sum_{j,\sigma_2} \mathcal{J}_{\sigma_2}{\hspace{-0.5mm}}\Vec{S}_{j,A}^{(1)} \cdot {\hspace{-0.5mm}}\Vec{S}^{(1)}_{j+\sigma_1,A}}_{\Hat{H}_{Ex,2}} \\
        & \underbrace{-\sum_{j} \Big[\mathcal{D}_x {\hspace{0.5mm}}S_{x,j,A}^{(1)2} + \mathcal{D}_y {\hspace{0.5mm}}S_{y,j,A}^{(1)2} + \mathcal{D}_z{\hspace{0.5mm}}S_{z,j,A}^{(1)2} \Big]}_{\Hat{H}_{An}} \\
        & \underbrace{-\sum_{j}\gamma \hbar \Vec{B}_0\cdot{\hspace{-0.5mm}}\Vec{S}^{(1)}_{j,A}}_{\Hat{H}_{Ext}},
    \end{aligned}
\end{equation}

\noindent
with
\begin{equation} \label{a2}
    \begin{aligned}
        J^{(1)}_{\Vec{\sigma}_{\pm a}}&=J^{(1)}_1, \\
        J^{(2)}_{\Vec{\sigma}_{\pm a}}&=J^{(2)}_1, \\
        J_{\Vec{\sigma}_{\pm d_1}} & =J_{\Vec{\sigma}_{\pm d_2}}=J_2, \\
        J^{(1)}_{\Vec{\sigma}_{\pm b}}&=J^{(1)}_3, \\
        J^{(2)}_{\Vec{\sigma}_{\pm b}}&=J^{(2)}_3, \\
    \end{aligned}
\end{equation}

\noindent
where \(\Vec{d}_1 = \frac{1}{2}(\Vec{a} + \Vec{b})\) and \(\Vec{d}_2 = \frac{1}{2}(\Vec{a} - \Vec{b})\), so that
\begin{equation} \label{a3}
    \begin{aligned}
        & \alpha^{(1)}_1 = 4(J^{(1)}_1 + J^{(1)}_3), \\
        & \beta^{(1)}_1 = 4(J^{(1)}_1\cos(k_xa) + J^{(1)}_3\cos(k_zb)), \\
        & \alpha^{(2)}_1 = 4(J^{(2)}_1 + J^{(2)}_3), \\
        & \beta^{(2)}_1 = 4(J^{(2)}_1\cos(k_xa) + J^{(2)}_3\cos(k_zb)), \\
        & \alpha_2 = 8J_2, \\
        & \beta_2 = 4J_2(\cos(\frac{1}{2}k_xa + \frac{1}{2}k_zb) + \cos(\frac{1}{2}k_xa - \frac{1}{2}k_zb).
    \end{aligned}
\end{equation}

\noindent
Here the superscript indicates the position of the local moment in the unit cell. The Hamiltonian for sublattice \(2\) is found by substitution of \(1 \rightleftharpoons 2\). The solutions now include in- and out-of-phase intralayer excitations. We list the LL matrices in the presence an external field along the \(b\)- and \(a\)-axes below.

\subsection{Ferromagnetic Phase - Easy Axis External Field}

\hspace{\parindent}The eigenvalue matrix in the basis of
\begin{equation} \label{a4}
    \begin{bmatrix}
        \hspace{-0.5mm}m^{(1)}_{x} \\
        \hspace{-0.5mm}m^{(2)}_{x} \\
        \hspace{-0.5mm}m^{(1)}_{y} \\
        \hspace{-0.5mm}m^{(2)}_{y},
    \end{bmatrix}\hspace{-1.2mm}.
\end{equation}

for an external field along the (positive) easy axis reads
\begin{align} \label{a5}
\begin{bmatrix}
0 & 0 & iA^{(1)} & -iA_1 \\
0 & 0 & -iA_1 & iA^{(2)} \\
-iB^{(1)}  & iB_1  & 0 & 0  \\
iB_1 & -iB^{(2)} & 0 & 0
\end{bmatrix}\hspace{-1.2mm},
\end{align}

\noindent
where
\begin{equation} \label{a6}
    \begin{aligned}
        A^{(1)} & = B_0 + 2(D_z - D_y) + [\alpha_1^{(1)} - \beta_1^{(1)}] + \alpha_2 + \mu_0 M_s f(k), \\
        A^{(2)} & = B_0 + 2(D_z - D_y) + [\alpha_1^{(2)} - \beta_1^{(2)}] + \alpha_2 + \mu_0 M_s f(k), \\
        B^{(1)} & = B_0 + 2(D_z - D_x) + [\alpha_1^{(1)} - \beta_1^{(1)}] \\
        & + \alpha_2 + \frac{k_x^2}{k^2}\mu_0 M_s(1-f(k)), \\
        B^{(2)} & = B_0 + 2(D_z - D_x) + [\alpha_1^{(2)} - \beta_1^{(2)}] \\
        & + \alpha_2 + \frac{k_x^2}{k^2}\mu_0 M_s(1-f(k)), \\
        A_1 & = \beta_2 - \mu_0 M_sf(k), \\
        B_1 & = \beta_2 - \frac{k_x^2}{k^2}\mu_0 M_s(1-f(k)).
    \end{aligned}
\end{equation}

\subsection{Canted Phase - Intermediate Axis External Field}

\hspace{\parindent}The eigenvalue matrix for the canted and saturated (\(\theta = \theta/2\) phases reads
\begin{align} \label{a7}
\begin{bmatrix}
0 & 0 & iC^{(1)} & -iC_1 \\
0 & 0 & -iC_1 & iC^{(2)} \\
-iD^{(1)}  & iD_1  & 0 & 0  \\
iD_1 & -iD^{(2)} & 0 & 0
\end{bmatrix}\hspace{-1.2mm},
\end{align}

\noindent
where
\begin{equation} \label{a8}
    \begin{aligned}
        C^{(1)} & = B_0\sin(\theta) + 2(D_x \sin^2(\theta) + D_z\cos^2(\theta) - D_y) \\
        & + [\alpha_1^{(1)} - \beta_1^{(1)}] + \alpha_2 + \mu_0 M_s f(k), \\
        C^{(2)} & = B_0\sin(\theta) + 2(D_x\sin^2(\theta) + D_z\cos^2(\theta) - D_y) \\
        & + [\alpha_1^{(2)} - \beta_1^{(2)}] + \alpha_2 + \mu_0 M_s f(k), \\
        D^{(1)} & = B_0\sin(\theta) + 2(D_z-D_x)\cos(2\theta) \\ 
        & + [\alpha_1^{(1)} - \beta_1^{(1)}] + \alpha_2 + \mu_0 M_s \frac{k_{\alpha}^2}{k^2}(1-f(k)), \\
        D^{(2)} & = B_0\sin(\theta) + 2(D_z - D_x)\cos(2\theta) \\ 
        & + [\alpha_1^{(2)} - \beta_1^{(2)}] + \alpha_2 + \mu_0 M_s \frac{k_{\alpha}^2}{k^2}(1-f(k)), \\
        C_1 & = \beta_2 - \mu_0 M_s f(k),\\
        D_1 & = \beta_2 - \mu_0 M_s \frac{k_{\alpha}^2}{k^2}(1-f(k)) .
    \end{aligned}
\end{equation}

\section{Two-sublattice Bilayer CrSBr} \label{appB}

The \(8\times8\) LL matrix for the bilayer with anisotropic exchange interactions cannot be analytically diagonalized but the listed forms here can be easily solved numerically. The Hamiltonian of the bilayer is the sum of the separate monolayers with an interaction term which now has two contributions in the form of \hyperref[4]{Eq. (4)}, one for each magnetic atom in the monolayer unit cell. The corresponding matrix forms of the LL equation for the AFM, FM and canted phases are given below.

\subsection{\label{app:subsec}Antiferromagnetic and ferromagnetic phase}

In the basis
\begin{equation} \label{b1}
    \begin{bmatrix}
        \hspace{-0.5mm}m^{(1)}_{x,A} \\
        \hspace{-0.5mm}m^{(1)}_{x,B} \\
        \hspace{-0.5mm}m^{(2)}_{x,A} \\
        \hspace{-0.5mm}m^{(2)}_{x,B} \\
        \hspace{-0.5mm}m^{(1)}_{y,A} \\
        \hspace{-0.5mm}m^{(1)}_{y,B} \\
        \hspace{-0.5mm}m^{(2)}_{y,A} \\
        \hspace{-0.5mm}m^{(2)}_{y,B} \\
    \end{bmatrix}\hspace{-1.2mm}.
\end{equation}

\noindent
the dynamical LL matrix for the AFM phase reads
\begin{align} \label{b2}
\setlength\arraycolsep{-0.9pt}
\begin{bmatrix}
0 & 0 & 0 & 0 & iE_1^{(1)} & -iJ_{\perp} & -iE & 0 \\
0 & 0 & 0 & 0 & iJ_{\perp} & -iE_2^{(1)} & 0 & iE \\
0 & 0 & 0 & 0 & -iE & 0 & iE_1^{(2)} & -iJ_{\perp} \\
0 & 0 & 0 & 0 & 0 & iE & iJ_{\perp} & -iE_2^{(2)} \\
-iF_1^{(1)}  & iJ_{\perp}  & iF  & 0 & 0 & 0 & 0 & 0  \\
-iJ_{\perp} & iF_2^{(1)} & 0  & -iF & 0 & 0 & 0 & 0 \\
iF  & 0  & -iF_1^{(2)}  & iJ_{\perp} & 0 & 0 & 0 & 0  \\
0 & -iF & -iJ_{\perp}  & iF_2^{(2)} & 0 & 0 & 0 & 0 \\
\end{bmatrix}\hspace{-1.2mm},
\end{align}

\noindent
where
\begin{equation} \label{b3}
    \begin{aligned}
        E_1^{(1)} & = B_0 + 2(D_z - D_y) + [\alpha_1^{(1)} - \beta_1^{(1)}]\\
        & + \alpha_2 - J_{\perp} + \mu_0 M_s f(k), \\
        E_1^{(2)} & = B_0 + 2(D_z - D_y) + [\alpha_1^{(2)} - \beta_1^{(2)}]\\
        & + \alpha_2 - J_{\perp} + \mu_0 M_s f(k), \\
        E_2^{(1)} & = -B_0 + 2(D_z - D_y) + [\alpha_1^{(1)}- \beta_1^{(1)}] \\
        &  + \alpha_2 - J_{\perp} + \mu_0 M_s f(k), \\
        E_2^{(2)} & = -B_0 + 2(D_z - D_y) + [\alpha_1^{(2)} - \beta_1^{(2)}] \\
        &  + \alpha_2 - J_{\perp} + \mu_0 M_s f(k), \\
        E & = \beta_2 - \mu_0 M_s f(k), \\
        F_1^{(1)} & = B_0 + 2(D_z - D_x) + [\alpha_1^{(1)} - \beta_1^{(1)}]\\
        & + \alpha_2 - J_{\perp} + \mu_0 M_s \frac{k_x^2}{k^2}(1-f(k)), \\
        F_1^{(2)} & = B_0 + 2(D_z - D_x) + [\alpha_1^{(2)} - \beta_1^{(2)}]\\
        & + \alpha_2 - J_{\perp} + \mu_0 M_s \frac{k_x^2}{k^2}(1-f(k)), \\
        F_2^{(1)} & = -B_0 + 2(D_z - D_x) + [\alpha_1^{(1)}- \beta_1^{(1)}]\\
        &  + \alpha_2 - J_{\perp} + \mu_0 M_s \frac{k_x^2}{k^2}(1-f(k)), \\
        F_2^{(2)} & = -B_0 + 2(D_z - D_x) + [\alpha_1^{(2)} - \beta_1^{(2)}] \\
        & + \alpha_2 - J_{\perp} + \mu_0 M_s \frac{k_x^2}{k^2}(1-f(k)), \\
        F & = \beta_2 - \mu_0 M_s \frac{k_x^2}{k^2}(1-f(k)). \\
    \end{aligned}
\end{equation}

\noindent
For the FM phase it reads
\begin{align} \label{b4}
\setlength\arraycolsep{-0.8pt}
\begin{bmatrix}
0 & 0 & 0 & 0 & iG^{(1)} & -iJ_{\perp} & -iG_1 & 0 \\
0 & 0 & 0 & 0 & -iJ_{\perp} & iG^{(1)} & 0 & -iG_1 \\
0 & 0 & 0 & 0 & -iG_1 & 0 & iG^{(2)} & -iJ_{\perp} \\
0 & 0 & 0 & 0 & 0 & -iG_1 & -iJ_{\perp} & iG^{(2)} \\
-iH^{(1)}  & iJ_{\perp}  & iH_1  & 0 & 0 & 0 & 0 & 0  \\
iJ_{\perp} & -iH^{(1)} & 0  & iH_1 & 0 & 0 & 0 & 0 \\
iH_1  & 0  & -iH^{(2)}  & iJ_{\perp} & 0 & 0 & 0 & 0  \\
0 & iH_1 & iJ_{\perp}  & -iH^{(2)} & 0 & 0 & 0 & 0 \\
\end{bmatrix}\hspace{-1.2mm},
\end{align}

\noindent
where
\begin{equation} \label{b5}
    \begin{aligned}
        G^{(1)} & = B_0 + 2(D_z - D_y) + [\alpha_1^{(1)} - \beta_1^{(1)}]\\
        & + \alpha_2 + J_{\perp} + \mu_0 M_s f(k), \\
        G^{(2)} & = B_0 + 2(D_z - D_y) + [\alpha_1^{(2)} - \beta_1^{(2)}]\\
        & + \alpha_2 + J_{\perp} + \mu_0 M_s f(k), \\
        G_1 & = \beta_2 - \mu_0 M_s f(k),\\
        H^{(1)} & = B_0 + 2(D_z - D_x) + [\alpha_1^{(1)}- \beta_1^{(1)}]\\
        &  + \alpha_2 + J_{\perp} + \mu_0 M_s \frac{k_x^2}{k^2}(1-f(k)), \\
        H^{(2)} & = B_0 + 2(D_z - D_x) + [\alpha_1^{(2)} - \beta_1^{(2)}] \\
        & + \alpha_2 + J_{\perp} + \mu_0 M_s \frac{k_x^2}{k^2}(1-f(k)), \\
        H_1 & = \beta_2 -  \mu_0 M_s \frac{k_x^2}{k^2}(1-f(k)).
    \end{aligned}
\end{equation}

\subsection{\label{app:subsec}Canted Phase}

When applying an external magnetic field along the intermediate axis, we expand the non-collinear phases  in the basis
\begin{equation} \label{b6}
    \begin{bmatrix}
        \hspace{-0.5mm}m^{(1)}_{\alpha,A} \\
        \hspace{-0.5mm}m^{(1)}_{\alpha,B} \\
        \hspace{-0.5mm}m^{(2)}_{\alpha,A} \\
        \hspace{-0.5mm}m^{(2)}_{\alpha,B} \\
        \hspace{-0.5mm}m^{(1)}_{\beta,A} \\
        \hspace{-0.5mm}m^{(1)}_{\beta,B} \\
        \hspace{-0.5mm}m^{(2)}_{\beta,A} \\
        \hspace{-0.5mm}m^{(2)}_{\beta,B} \\
    \end{bmatrix}\hspace{-1.2mm}.
\end{equation}

\noindent
For both canted and saturated (\(\theta = \theta/2\)) spin textures
\begin{align} \label{b7}
\setlength\arraycolsep{-1.1pt}
\begin{bmatrix}
0 & 0 & 0 & 0 & iI^{(1)} & -iJ_{\perp} & -iI_1 & 0 \\
0 & 0 & 0 & 0 & -iJ_{\perp} & iI^{(1)} & 0 & -iI_1 \\
0 & 0 & 0 & 0 & -iI_1 & 0 & iI^{(2)} & -iJ_{\perp} \\
0 & 0 & 0 & 0 & 0 & -iI_1 & -iJ_{\perp} & iI^{(2)} \\
-iK_A^{(1)}  & iK_2  & iK_{1,A}  & 0 & 0 & 0 & 0 & 0  \\
iK_2 & -iK_B^{(1)} & 0  & iK_{1,B} & 0 & 0 & 0 & 0 \\
iK_{1,A}  & 0  & -iK_A^{(2)}  & iK_2 & 0 & 0 & 0 & 0  \\
0 & iK_{1,B} & iK_2  & -iK_B^{(2)} & 0 & 0 & 0 & 0 \\
\end{bmatrix}\hspace{-1.2mm},
\end{align}

\noindent
where
\begin{equation} \label{b8}
    \begin{aligned}
        I^{(1)} & = B_0\sin(\theta) + 2(D_x\sin^2(\theta) + D_z\cos^2(\theta) - D_y) \\
        & + [\alpha_1^{(1)} - \beta_1^{(1)}] + \alpha_2 - J_{\perp}\cos(2\theta) + \mu_0 M_s f(k),\\
        I^{(2)} & = B_0\sin(\theta) + 2(D_x\sin^2(\theta) + D_z\cos^2(\theta) - D_y) \\
        & + [\alpha_1^{(2)} - \beta_1^{(2)}] + \alpha_2 - J_{\perp}\cos(2\theta) + \mu_0 M_s f(k), \\
        I_1 & = \beta_2 -\mu_0 M_s f(k),\\
        K_A^{(1)} & = B_0\sin(\theta) + 2(D_z - D_x)\cos(2\theta) + [\alpha_1^{(1)} - \beta_1^{(1)}]\\
        & + \alpha_2 - J_{\perp}\cos(2\theta)+ \mu_0 M_s \frac{k_{\alpha,A}^2}{k^2}(1-f(k)), \\
        K_B^{(1)} & = B_0\sin(\theta) + 2(D_z - D_x)\cos(2\theta) + [\alpha_1^{(1)} - \beta_1^{(1)}]\\
        & + \alpha_2 - J_{\perp}\cos(2\theta)+ \mu_0 M_s \frac{k_{\alpha,B}^2}{k^2}(1-f(k)), \\
        K_A^{(2)} & = B_0\sin(\theta) + 2(D_z - D_x)\cos(2\theta) + [\alpha_1^{(2)} - \beta_1^{(2)}]\\
        & + \alpha_2 - J_{\perp}\cos(2\theta)+ \mu_0 M_s \frac{k_{\alpha,A}^2}{k^2}(1-f(k)), \\
        K_B^{(2)} & = B_0\sin(\theta) + 2(D_z - D_x)\cos(2\theta) + [\alpha_1^{(2)} - \beta_1^{(2)}]\\
        & + \alpha_2 - J_{\perp}\cos(2\theta)+ \mu_0 M_s \frac{k_{\alpha,B}^2}{k^2}(1-f(k)), \\
        K_{1,A} & = \beta_2 - \mu_0 M_s \frac{k_{\alpha,A}^2}{k^2}(1-f(k)),\\
        K_{1,B} & = \beta_2 - \mu_0 M_s \frac{k_{\alpha,B}^2}{k^2}(1-f(k)),\\
        K_2 & = -J_{\perp}\cos(2\theta). \\
    \end{aligned}
\end{equation}

In the limit of isotropic interlayer exchange \(J_1^{(1)} = J_1^{(2)}\) and \(J_3^{(1)} = J_3^{(2)}\), the solutions for a single atom per sublattice in the text correspond to the intralayer in-phase excitations, i.e.
\begin{equation} \label{b9}
    \begin{aligned}
        & \hspace{-0.5mm}m^{(1)}_{\alpha,A} = \hspace{-0.5mm}m^{(2)}_{\alpha,A}, \\
        & \hspace{-0.5mm}m^{(1)}_{\beta,A} = \hspace{-0.5mm}m^{(2)}_{\beta,A}, \\
        & \hspace{-0.5mm}m^{(1)}_{\alpha,B} = \hspace{-0.5mm}m^{(2)}_{\alpha,B}, \\
        & \hspace{-0.5mm}m^{(1)}_{\beta,B} = \hspace{-0.5mm}m^{(2)}_{\beta,B}, \\
    \end{aligned}
\end{equation}

\noindent
but the dipolar field is twice as large by effective doubling of lattice sites. The intralayer out-of-phase excitations, or
\begin{equation} \label{b10}
    \begin{aligned}
        & \hspace{-0.5mm}m^{(1)}_{\alpha,A} = -\hspace{-0.5mm}m^{(2)}_{\alpha,A}, \\
        & \hspace{-0.5mm}m^{(1)}_{\beta,A} = -\hspace{-0.5mm}m^{(2)}_{\beta,A}, \\
        & \hspace{-0.5mm}m^{(1)}_{\alpha,B}e = -\hspace{-0.5mm}m^{(2)}_{\alpha,B}, \\
        & \hspace{-0.5mm}m^{(1)}_{\beta,B} = -\hspace{-0.5mm}m^{(2)}_{\beta,B}, \\
    \end{aligned}
\end{equation}

\noindent
correspond to the higher (optical) band folded back in the reduced Brillouin zone. The intralayer in-phase (out-of-phase) modes are characterized by parameter combinations \(\alpha_i - \beta_i\) (\(\alpha_i + \beta_i\)) \cite{strain}.

\section{Coordinate transformation and base vectors for non-collinear spin textures} \label{appC}

In the following Section we describe the coordinate transformations that simplify the solution of the LL equation when the spin texture is non-collinear.
\vspace{-5mm}
\subsection{\label{app:subsec}Monolayer - Canted Phase}

\begin{figure}[h] \label{fig5}
    \centering
    \includegraphics[scale=0.35]{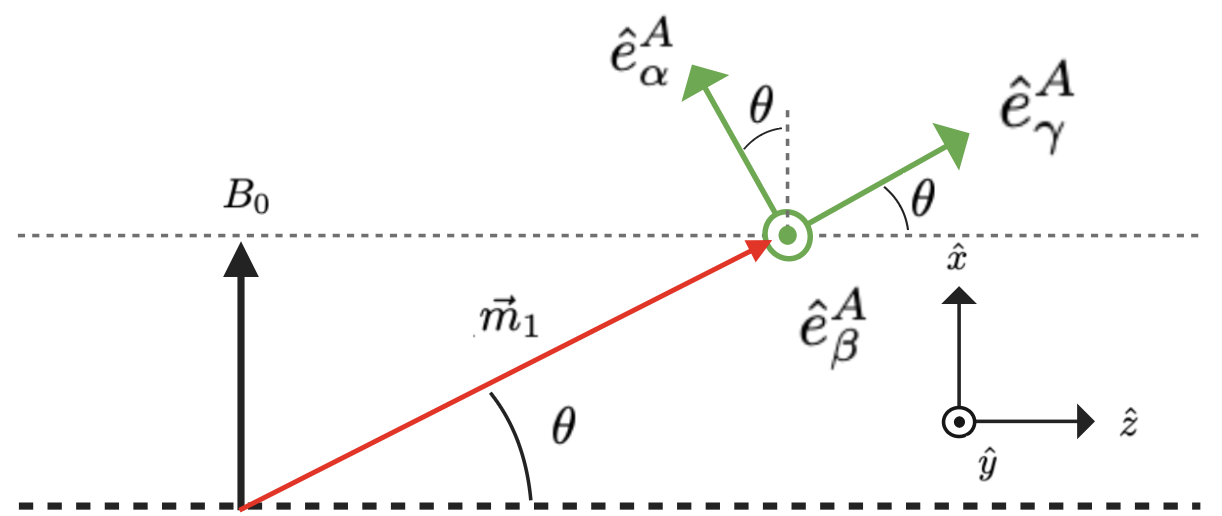}
    \caption{Canted orientation of magnetic moments in the monolayer when \( \Vec{B}_{Ext}\parallel \Hat{x}\), including unit vectors in the rotated basis.}
\end{figure}

The canted phase in the monolayer is described in the basis of \([\Hat{e}^A_{\alpha},\Hat{e}^A_{\beta},\Hat{e}^A_{\gamma}]\) with transformations 
\begin{equation} \label{c1}
    \begin{aligned}
        & \Hat{x} =  \Hat{e}^A_{\alpha}\cos \theta  + \Hat{e}^A_{\gamma} \sin \theta , \\
        & \Hat{y} = \Hat{e}^A_{\beta}, \\
        & \Hat{z} = -\sin(\theta)\Hat{e}^A_{\alpha} + \cos(\theta)\Hat{e}^A_{\gamma},
    \end{aligned}
\end{equation}

\noindent
as illustrated in \hyperref[fig5]{FIG. 5}.

\vspace{-10mm}
\subsection{\label{app:subsec}Bilayer - Canted Phase}

\begin{figure}[h!] \label{fig6}
    \centering
    \includegraphics[scale=0.27]{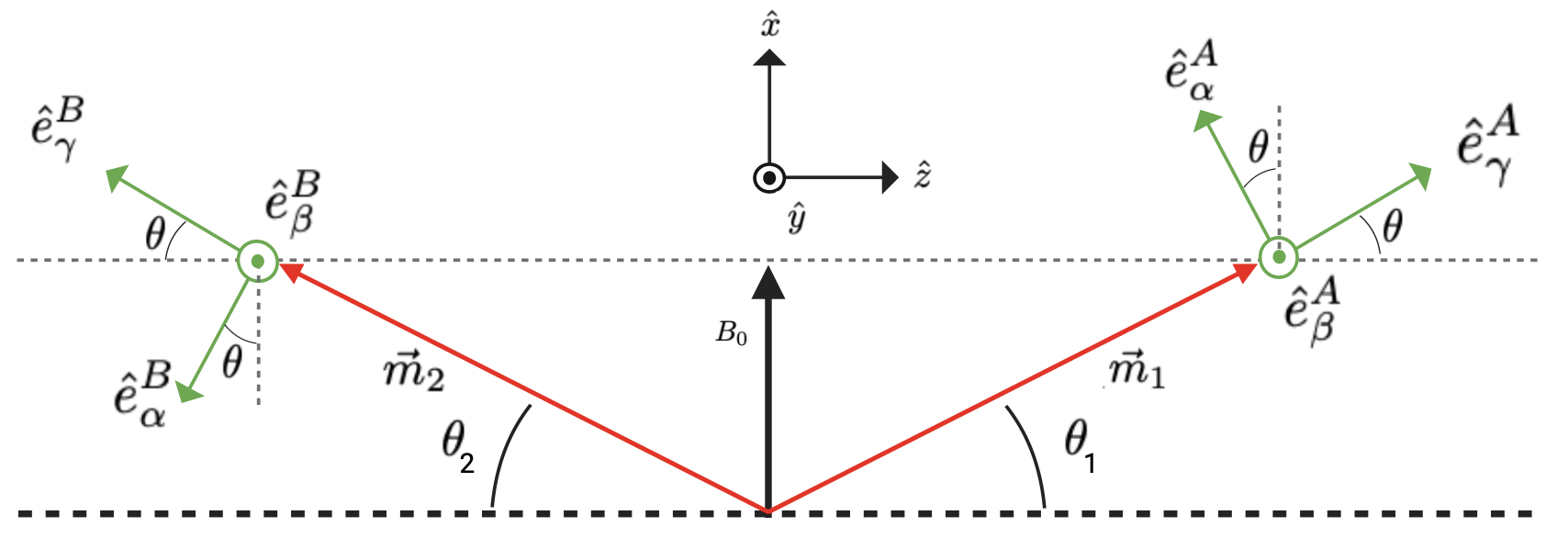}
    \caption{Orientation of magnetic moments in the bilayer with \( \Vec{B}_{Ext}\parallel \Hat{x}\), including unit vectors in non-colinear basis. \(A\) and \(B\) indicate magnetic moments in layer \(A\) and \(B\), respectively.}
\end{figure}

The canted phase in the bilayer is described in the bases of \([\Hat{e}^A_{\alpha},\Hat{e}^A_{\beta},\Hat{e}^A_{\gamma}]\) and \([\Hat{e}^B_{\alpha},\Hat{e}^B_{\beta},\Hat{e}^B_{\gamma}]\) with transformations
\begin{equation} \label{c2}
    \begin{aligned}
        & \Hat{x} = \cos(\theta)\Hat{e}^A_{\alpha} + \sin(\theta)\Hat{e}^A_{\gamma}, \\
        & \Hat{y} = \Hat{e}^A_{\beta}, \\
        & \Hat{z} = -\sin(\theta)\Hat{e}^A_{\alpha} + \cos(\theta)\Hat{e}^A_{\gamma},
    \end{aligned}
\end{equation}

\noindent
and
\begin{equation} \label{c3}
    \begin{aligned}
        & \Hat{x} = -\cos(\theta)\Hat{e}^B_{\alpha} + \sin(\theta)\Hat{e}^B_{\gamma}, \\
        & \Hat{y} = \Hat{e}^B_{\beta}, \\
        & \Hat{z} = -\sin(\theta)\Hat{e}^B_{\alpha} - \cos(\theta)\Hat{e}^B_{\gamma},
    \end{aligned}
\end{equation}

\noindent
as illustrated in \hyperref[fig6]{FIG. 6}.

\vspace{100mm}

\section{Bilayer Phase Transitions} \label{appD}

In this Section, we address the spin-flip and spin canting phase transitions illustrated.

\subsection{Spin-flip Transition}

\begin{figure}[h] \label{fig7}
    \centering
    \includegraphics[scale=0.22]{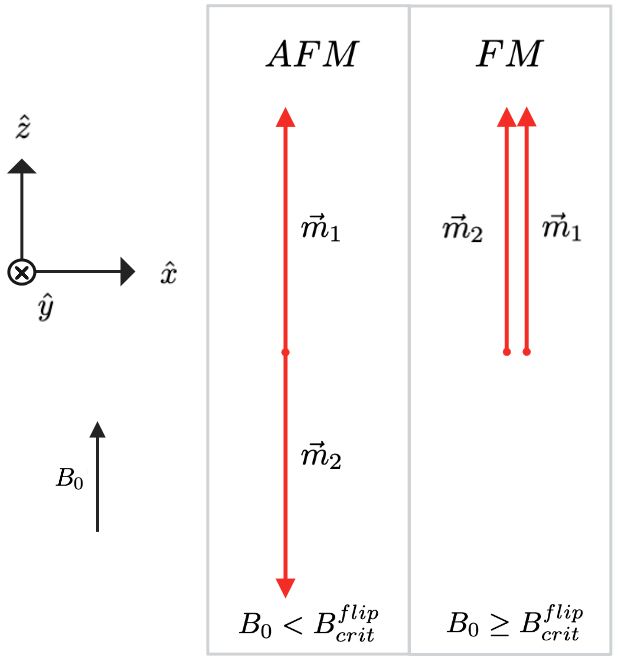}
    \caption{Spin-flip phase transition in the bilayer.}
\end{figure}

A ``spin-flip" transition occurs at external fields  \(B_0 = B^{crit}_{flip}\) along the easy axis (\(z\)) as illustrated in \hyperref[fig7]{FIG. 7}. It should not be confused with the spin-flop transition in materials with stronger interlayer exchange coupling.

\vspace{-8mm}
\subsection{Canted Phase Transition}

\begin{figure}[h] \label{fig8}
    \centering
    \includegraphics[scale=0.22]{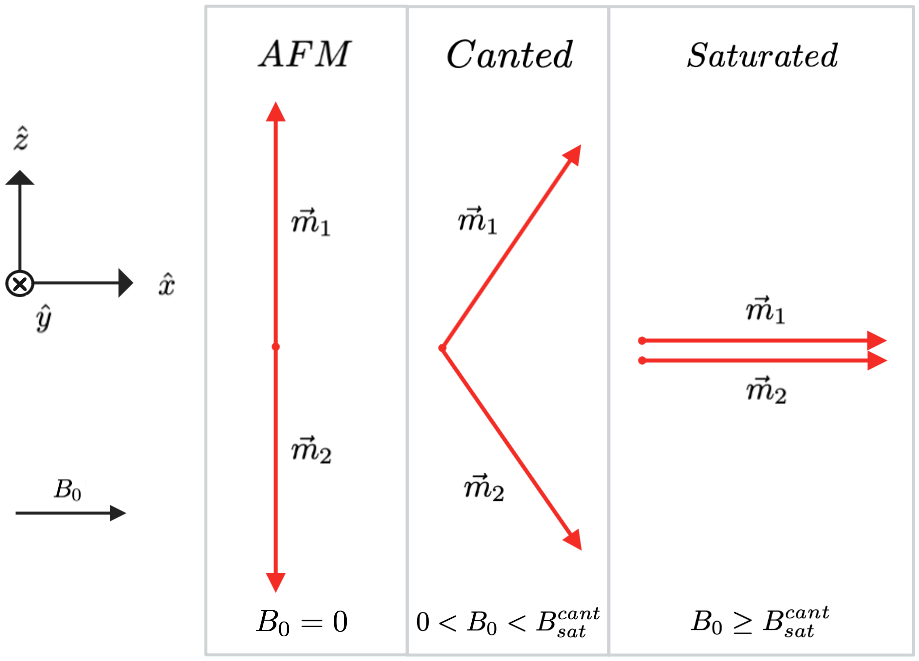}
    \caption{Canted phase transition in the bilayer.}
\end{figure}

The continuous canting transition occurs for external fields oriented along the intermediate axis (\(x\)) in the bilayer (monolayer) for any field \(B < B_0 < B^{cant}_{sat}\) (\(0 < B_0 < B_{sat}\)). The magnetic moments are parallel to the intermediate axis for higher fields as illustrated in \hyperref[fig8]{FIG. 8}.

\vspace{15mm}
\bibliography{apssamp}

\end{document}